Note: This is the final draft of a technical report from the Computing Community Consortium of the Computing Research Association. The published version is referenced as P. Michelucci, L. Shanley, J. Dickinson, and H. Hirsh, "A U.S. Research Roadmap for Human Computation," Computing Community Consortium Technical Report, 2015.
and is available at: http://www.cra.org/ccc/files/docs/meetings/HumanComputation/Final%20HC%20Report.pdf.

# A U.S. Research Roadmap for Human Computation

## A COMMUNITY REPORT FROM THE 2014 HUMAN COMPUTATION ROADMAP SUMMIT

CCC
Computing Community Consortium
Catalyst

# A U.S. Research Roadmap for Human Computation

**A Community Report from the 2014 Human Computation Roadmap Summit**

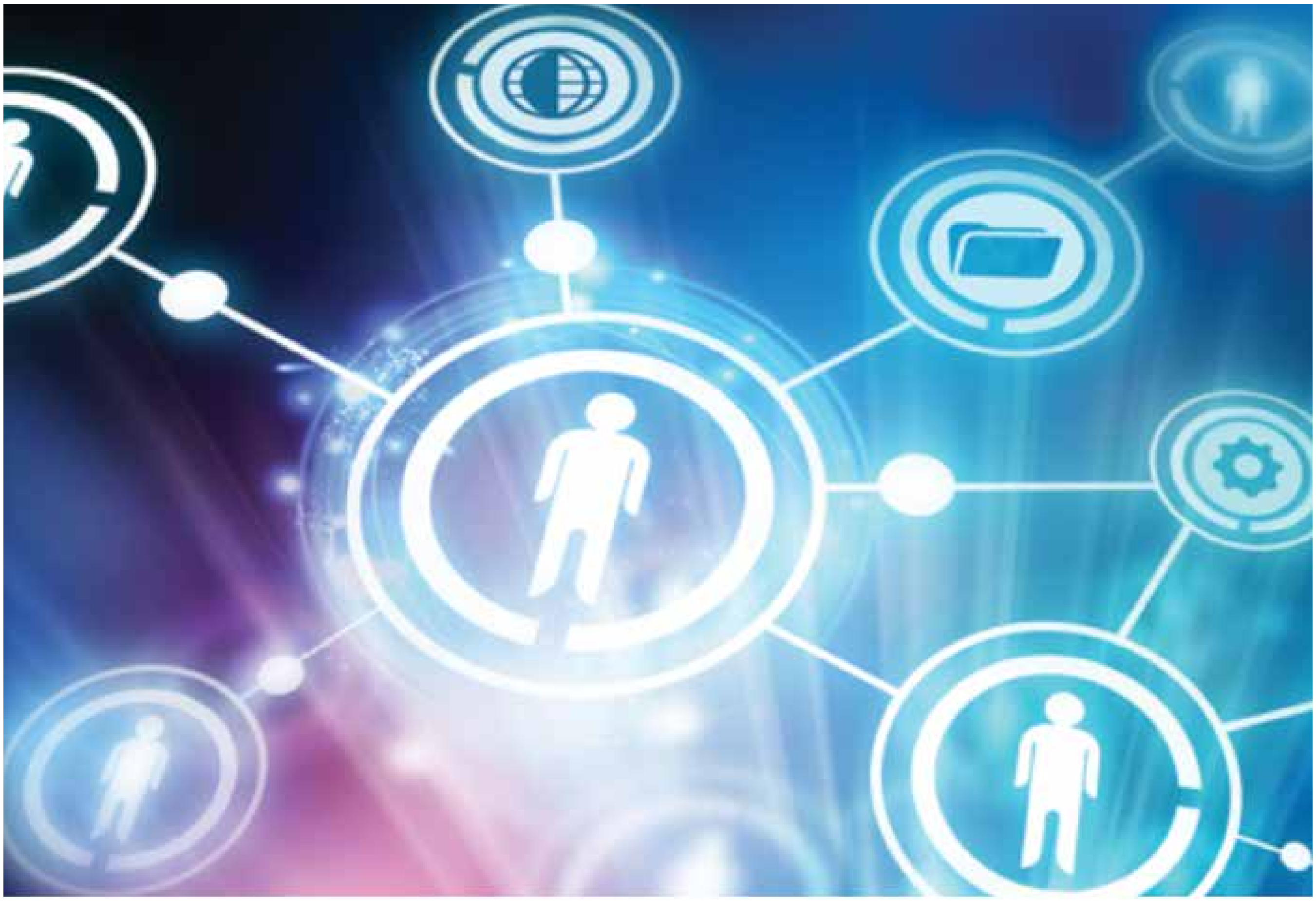

**Organized by**

Human Computation Institute

Woodrow Wilson International Center for Scholars

Cornell University

**Sponsored by**

CCC
Computing Community Consortium
Catalyst

CRA
Computing Research Association

## Summary

The Web has made it possible to harness human cognition en masse to achieve new capabilities. Some of these successes are well known; for example Wikipedia has become the go-to place for basic information on all things; Duolingo engages millions of people in real-life translation of text, while simultaneously teaching them to speak foreign languages; and fold.it has enabled public-driven scientific discoveries by recasting complex biomedical challenges into popular online puzzle games. These and other early successes hint at the tremendous potential for future crowd-powered capabilities for the benefit of health, education, science, and society. In the process, a new field called *Human Computation* has emerged to better understand, replicate, and improve upon these successes through scientific research.

Human Computation refers to the science that underlies online crowd-powered systems and was the topic of a recent visioning activity in which a representative cross-section of researchers, industry practitioners, visionaries, funding agency representatives, and policy makers came together to understand what makes crowd-powered systems successful. Teams of experts considered past, present, and future human computation systems to explore which kinds of crowd-powered systems have the greatest potential for societal impact and which kinds of research will best enable the efficient development of new crowd-powered systems to achieve this impact. In this report, we summarize the products and findings of those activities. The unconventional process and activities employed by the workshop were informed by human computation research and are described in the Appendix.

## Humans and Machines Participating in the Same System

Solving today's most challenging and complex problems requires an ability to build consensus around common goals and gather, process, and act on information at massive scales with increasing efficiency. We do not know how to create machines with the critical cognitive abilities required for solving important human-centered problems. But what if we could engineer systems that combine the respective strengths of machines and humans toward new capabilities?

Human Computation is an emerging field that considers the design and analysis of information processing systems in which humans participate as computational agents[1]. A multidisciplinary community of academics, visionaries, private industry researchers, and federal program officers, met to explore the transformative potential of directly employing human cognition within larger computational systems. During a three-day workshop held in June 2014, we explored the full landscape of human computation. We considered stakeholders and goals, examined historical successes, designed promising new systems, and ultimately sought to identify high-impact research strategies for achieving near-term societal benefits.

## Impact of Human Computation

We found that human computation methods have stimulated the economy via an online workforce ecosystem[2], which includes crowdsourced labor markets[3], contributive vocational training[4], innovation crowdfunding[5], and microlending to third-world entrepreneurs[6]. Human computation also has been used to support important behavioral change (e.g., to encourage health-related behaviors) via social networks[7][8], accelerate research[9], educate the public[10] through citizen science, enable new modes of civic engagement[11] through democratic processes, and reduce geopolitical conflict[12] through participatory gaming. It has been used to crowdsource the world's most comprehensive encyclopedia via massively distributed contributions and sharing of knowledge. And when extended into the physical world via participatory sensing (i.e., geographically distributed data acquisition and sharing using mobile devices or sensors) and coordinated action, human computation methods have been employed to save lives by amplifying situational awareness and coordinating rescue actions for crisis relief[13]. Furthermore, emergent human computation has been used to improve real-time epidemiology via predictive analytics and to reliably anticipate world events via social informatics[14].

## A Success Case

Notable successes, such as the fold.it project (see ***Figure 1***), have demonstrated dramatic results[15] using even simple human computation project designs. Fold.it is an online puzzle game that has enticed thousands of Internet participants to contribute their mental energies to folding virtual proteins. Recasting a biomedical research activity into a game that doesn't require specialized medical knowledge enabled thousands of volunteers to solve a problem and helped researchers better understand protein structures. In only a few weeks' time, it gave rise to the discovery of the tertiary structure of a regulatory protein for the pro-simian immunodeficiency virus (SIV), which previously eluded the research community for decades[16], and now may lead to new medications to treat the AIDS virus. At the time of this writing, fold.it is generating promising molecular topologies that could lead to treatment targets for the Ebola virus[17].

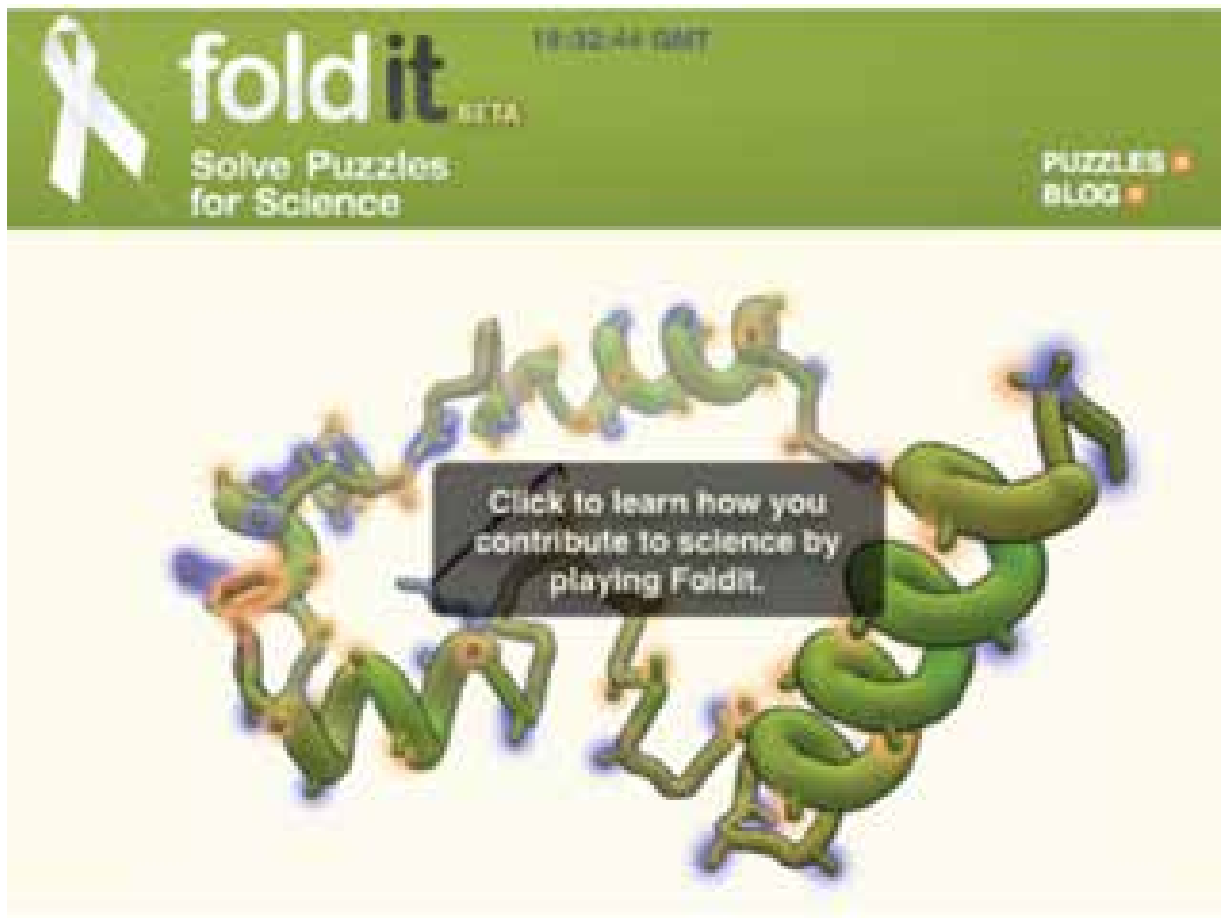

*Figure 1. The fold.it online protein folding game incentivized tens of thousands of citizens to contribute their intellectual resources to curing diseases.*

## Toward Repeatability

Efforts to replicate such successes have led to mixed results. This stems from the difficulty of ascertaining the precise, complex, and multidisciplinary combination of ingredients necessary for effective sustainable human computation. In traditional computer science, machine-based systems tend to exhibit predictable behavior such that machine errors can almost always be traced to faulty instructions. Indeed, a mature body of theoretical and applied methods exists today as a result of decades of funded research to support the development of such deterministic systems. Due to the vagaries of human behavior, however, these traditional methodologies are inadequate for human computation[18], which suggests the need for a new approach. The next section describes results from a set of ideation exercises to help inform new classes of methods better suited to systems with humans in the loop.

## New Project Ideas

In order to explore in depth the unique set of considerations that arise in human computation, we formed multidisciplinary breakout teams to develop new human computation project ideas. In addition to suggesting promising new potential capabilities, this exercise helped reveal risks and opportunities specific to this field and illuminated critical research areas that will be instrumental in advancing the field. Six projects are described below. When setting up the breakout groups, the organizers gave permission for one group to "do evil", that is, to formulate ways that human computation systems could be manipulated to yield outcomes that are socially undesirable. This is the last project idea described on page 10.

### Project Houston

On April 14, 1970, 56 hours into its space mission to the moon, Apollo 13 transmitted, "Houston, we've had a problem." What followed was a calmly heroic effort of remote engineering that led to a safe return. Imagine if we all could call on a calm, competent voice when we really needed help. We envision this resource as Project Houston, a mixed computer/human-computation service for distressed people. Overwhelming situations cause enormous societal and economic costs: violence, suicide attempts, hunger, lack of transport, homelessness, failure to access care, and job loss. When unaddressed, these problems tend to intensify, piling misery upon misery. They especially touch those least equipped to solve their problems including: the elderly infirm, people with mental illnesses, the isolated, ex-prisoners and youth. Individuals do not always have the appropriate expertise to address these problems but could benefit from access to the expertise of others. For example, the Apollo 13 mission landed successfully because they were able to access remote expertise in Houston and used their knowledge to redesign the spacecraft's carbon dioxide scrubbers.

Specialist social-work and mental-health workers have neither the numbers nor the 24-hour availability needed to aid people in distress before problems get worse. Project Houston would address these issues by using

### Project Houston

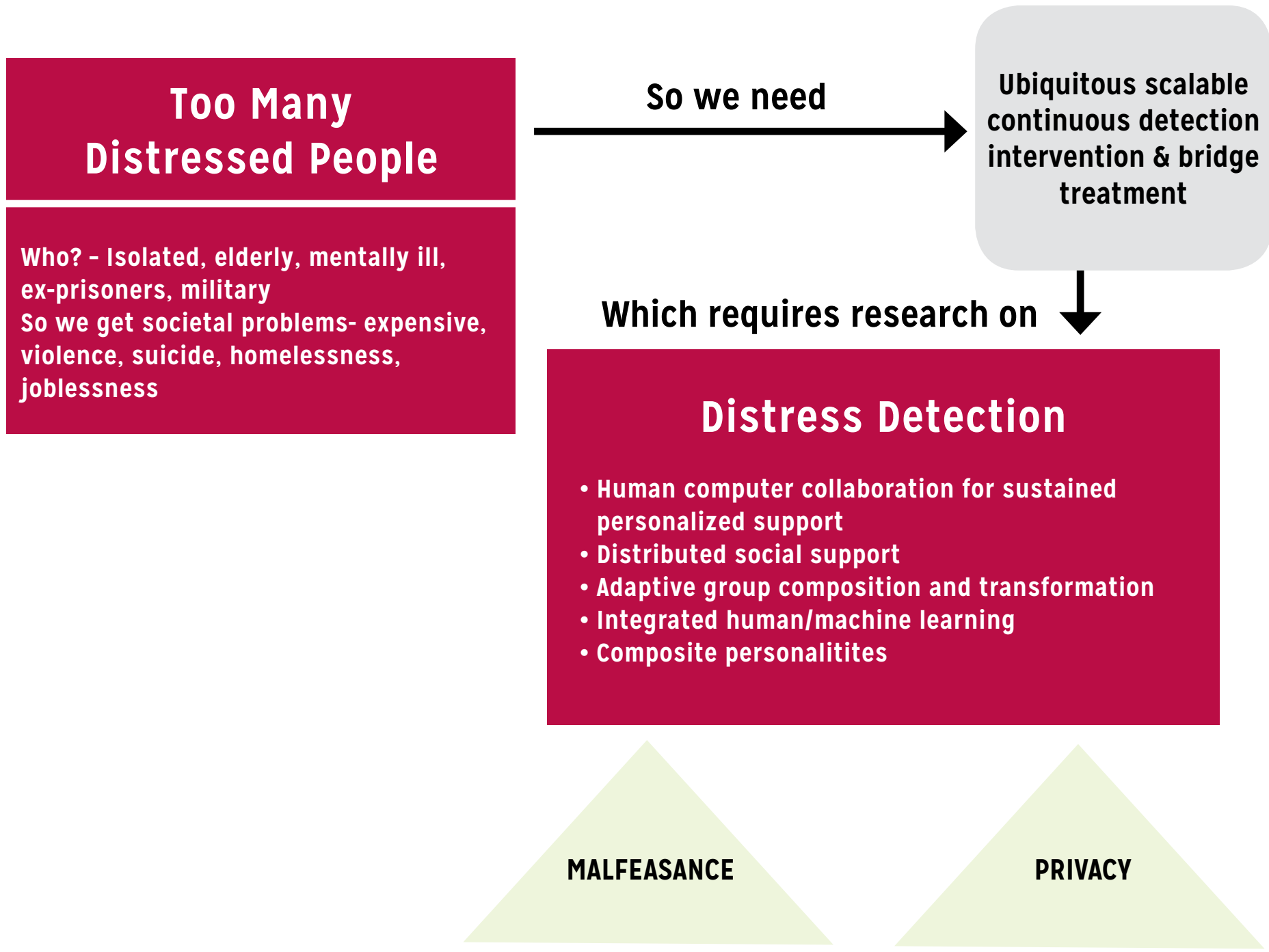

*Figure 2: Project Houston project diagram - 24-hour distributed assistance for people in distress.*

state of the art sensing, speech analysis, and natural language understanding to detect distress and offer help. Once help is requested, it would provide triage and first-level care using crowd-sourced, computer-supported composite personalities, bringing together the various traits needed to support the person. By semi-automatically assembling dynamic teams of volunteers, along with low and high level specialists, Project Houston could provide immediate 24-hour assistance. With computer-supported persistent memory and response integration enhanced by continuous machine learning, Project Houston could provide a consistently kind and patient personality even if the "crowd" changes completely over time in response to an escalating problem until the problem is resolved – just like mission control.

### Pathways to Radiology

With soaring numbers of individuals who are unemployed and under employed, the global community needs novel approaches to train workers and transition them into fulfilling professions. Crowdsourcing is a rapidly growing sector of the online economy where workers around the world perform tasks of short duration for small monetary incentives. Through online crowds, employers have access to a highly scalable, sometimes largely unskilled workforce. Currently, crowdsourcing typically only leverages basic cognitive abilities and may not train people with skills that could transfer to professional settings. In concert with this shift, recent advances in massively open online courses provide unprecedented opportunities for skills training, which when combined with performance of online tasks, could lead to measurable enhancement of skills needed in the offline workforce.

We explored a vision to combine online education with crowdsourced work in a way that provides pathways between low-skill micro-task crowdsourcing and the more complex tasks associated with professional vocations. The web app "Duolingo," is an example of this vision. It offers free language lessons while simultaneously creating value as a document translation service. If this dual-purpose strategy pays dividends, why stop with language learning and translation? The right kind of online ecosystem can enable endless opportunities to "level up" and learn new on-the-job skills, while also creating real value in the labor marketplace. Each worker could perform tasks, learn new skills, and receive credentials that enable them to take on more complex roles – such as reviewing and training others.

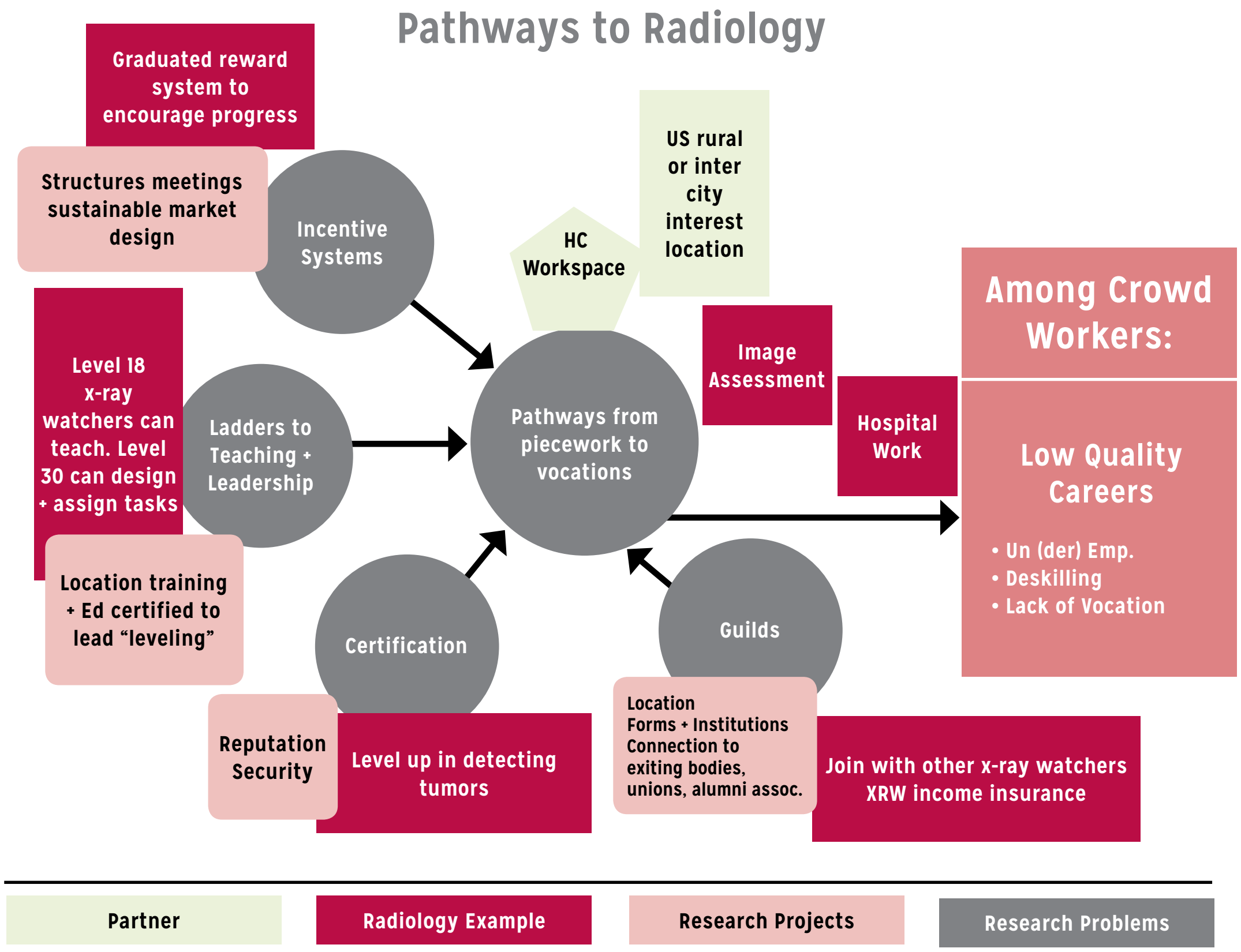

*Figure 3: Pathways to Radiology project diagram – crowdsourcing analysis while building a skilled labor force.*

This could be used in the field of radiology. Novice workers may enter the task market to perform tumor detection on x-ray images; this object recognition task is difficult for vision algorithms to perform reliably, but is a natural fit for crowdsourcing. As workers demonstrate proficiency, they may graduate to judge more difficult films by looking at edge cases that have less overall agreement, and then move on to write training materials for future workers, or review performance for a staff of newbies. Within this progression, the worker has learned about the subtleties of tumor detection and helped to author materials that propagate this knowledge throughout the system. We believe that online learning that doubles as work (and vice versa) can have a transformative impact on the future of work and education.

### Optimizing Effective Utilization of Social Services

On average, the poorest 20% of American families earn only $7,600 before taxes - approximately half of the federal poverty line for a 2-person household[19]. In addition to the obvious relationship between poverty and the difficulty meeting basic needs such as food, clothing and shelter, negative effects ripple out into other areas of well-being such as education, domestic abuse, and mental and physical health. While many state and federal programs exist to try to address issues related to poverty through social welfare programs, a family's burden of accessing those appropriate programs is often prohibitively high. At present, if a family is eligible for multiple services, they must be aware that the service exists, and then make and keep individual intake appointments with each potential service provider. This is grossly inefficient. Indeed, navigating the existing system represents a disproportionate hardship for marginalized populations such as homeless individuals, people with disabilities, and the working poor, groups that are most likely to benefit from these very programs.

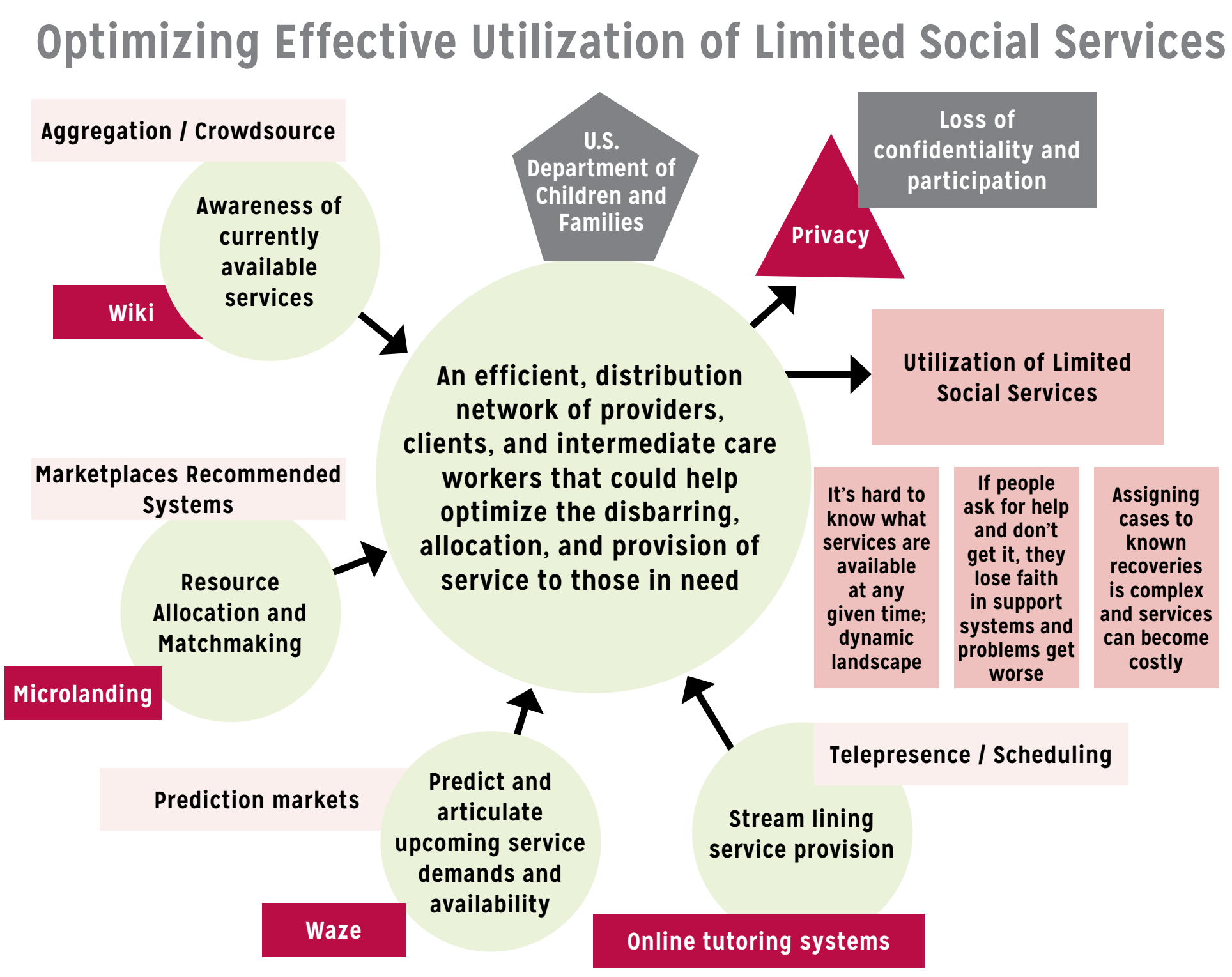

*Figure 4: Optimizing Effective Utilization of Limited Social Services project diagram*

Imagine if instead of requiring those in need to learn about each individual service and then travel to individual offices to check their eligibility, we were to harness the power of communities to solve this problem. Innovation in the area of human computation could make this possible. First, crowdsourcing platforms could be leveraged to aggregate and centralize information about all social welfare programs. This would include information such as eligibility requirements and real-time availability of the service (i.e. number of beds available in a drug treatment program and wait lists for mental health services.). The next step would be to streamline the process of verifying eligibility and bringing the process out of city hall and back into the community by empowering community members to serve as liaisons. If liaisons have full access to the centralized system and the ability to facilitate the enrollment process, a family no longer has to provide information multiple times and would become aware of the broad range of services available in their area. Through the use of both virtual and in-person human processing power, the solution to this resource allocation problem may well be within our grasp.

In just this past year, we have seen the dramatic impact of a similar technical challenge of health insurance enrollment through the Affordable Care Act. However, as evidenced by the frustrating roll-out of healthcare marketplaces, it is clear that the problem of coordination across multiple disparate agencies is nontrivial. Additionally, the process of verifying eligibility for social services is somewhat more complex than eligibility for health insurance subsidies, which are based solely on income. Because of these additional challenges, the intentional and efficient incorporation of human intervention is a necessary component of any successful solution. The benefits to streamlining access to social welfare programs could be enormous. With nearly $1 trillion spent by state and federal governments to fund social welfare programs, we can stand to gain in efficiency by streamlining information flow about access to these resources.

### Predicting Technology Trends

New technology is often a disruptive economic force, because it is hard to understand and can be enormously hyped. The resultant market volatility creates great upheaval in the economy and in workers' lives. To fix problems such as these, and help avoid financial crises due to overinvestment in technology startups, we need to be able to better estimate when is the right time to invest in technology.

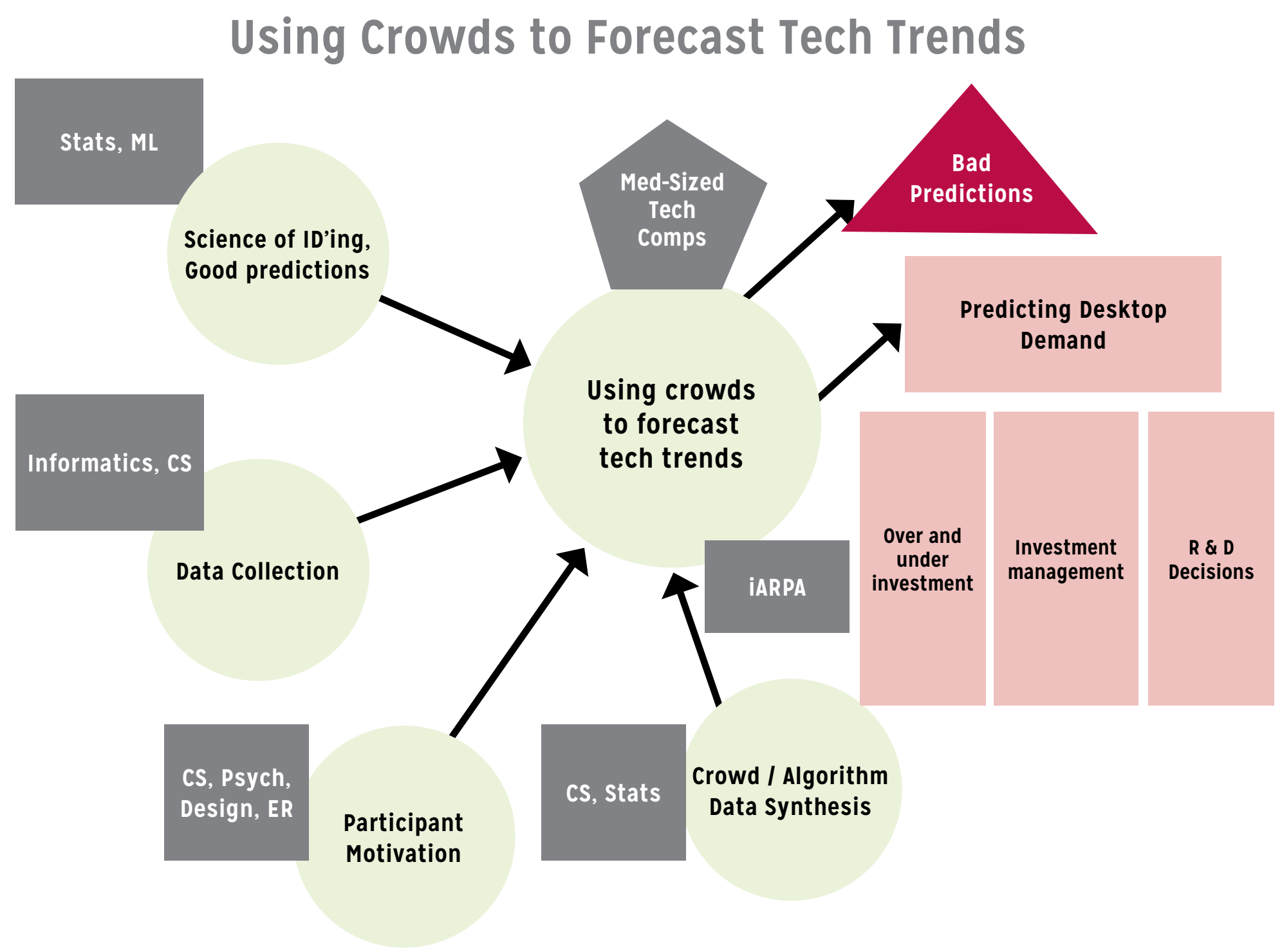

*Figure 5: Predicting Technology Trends project diagram – hybrid predictive models for better forecasting*

Today, tech news travels through news sites and is analyzed by domain experts but not necessarily technology experts. For example, in 2012, there were expectations about a meteoric rise in Massive Open Online Courses (MOOCs). There were fears that they would upset traditional universities by the over investment of money and time in their development, such as San Jose State University's heavily publicized but failed efforts [21]. The idea was good, but it was not the right time for heavy investment. To fix problems such as these, we need to be able to better estimate when is the right time to invest in technology.

Although knowing the future with certainty is of course impossible, it is possible to make much more informed forecasts of technological impact – driven by data, research, and expertise in the dynamics of technological growth. For example, IARPA (Intelligence Advanced Research Projects Activity) in the Office of the Director of National Intelligence has been successfully predicting political events in efforts such as The Good Judgment Project. Meeting participants proposed to do the same for technology foresight, using technology experts and models of growth in the technology sector.

The crowdsourcing research challenges involve identifying the right experts in the crowd and building interactive machine learning models that can be trained and improved over time. Right now, only people with the right knowledge can inform these models, and these people are spread across the world. We envision developing a platform that will bring together and organize this knowledge from experts. The collective knowledge would be used to develop models with much better predictive accuracy. Access to this information will ensure steady technological investing and provide protection against devastating tech bubbles.

### UpRiver

Around the world, humanitarian teams are constantly responding to devastation caused by extreme events. For example, exceptional rainfall upstream can wreak havoc on those living in the floodplains. There are many attempts to design and implement early warning systems, but too often vulnerable communities do not access, understand, or trust the information produced by others.

The Red Cross Red Crescent Climate Centre, in collaboration with the Engagement Lab at Emerson College and partners in developing countries, is developing the pervasive game "UpRiver," which extends into the real world and aims to:

- Improve river level data collection (and thus hydrological models to predict floods)
- Increase chances of communities trusting and acting on early warnings

UpRiver utilizes *people* as *sensors*, that is, players observe real-world river levels and report the levels via text message service. Players can also submit their 'forecast river level' (a guess) with a certain lead time (for example 48 hours). Whoever submits the forecast that is closest to the observed value wins an imaginary point. Players can use data from upstream communities to try to improve their forecasts.

Eventually when a good-enough predictive hydrological model is developed, the model will be added as a player ("Mike"). Participants who submit their forecast before the deadline (for example 8am) will receive a text message one minute after the deadline, indicating Mike's submission. Players that perform better than the model also earn a point. This will help people notice that the model tends to be accurate and trust the information. For example, if river levels are rising, and Mike predicts the river level to be ~3 meters above their home's kitchen floor, they would be more likely to act on that information. Eventually the trust earned through gameplay should help communities take the early warning seriously. More information about this evolving initiative is available online[1].

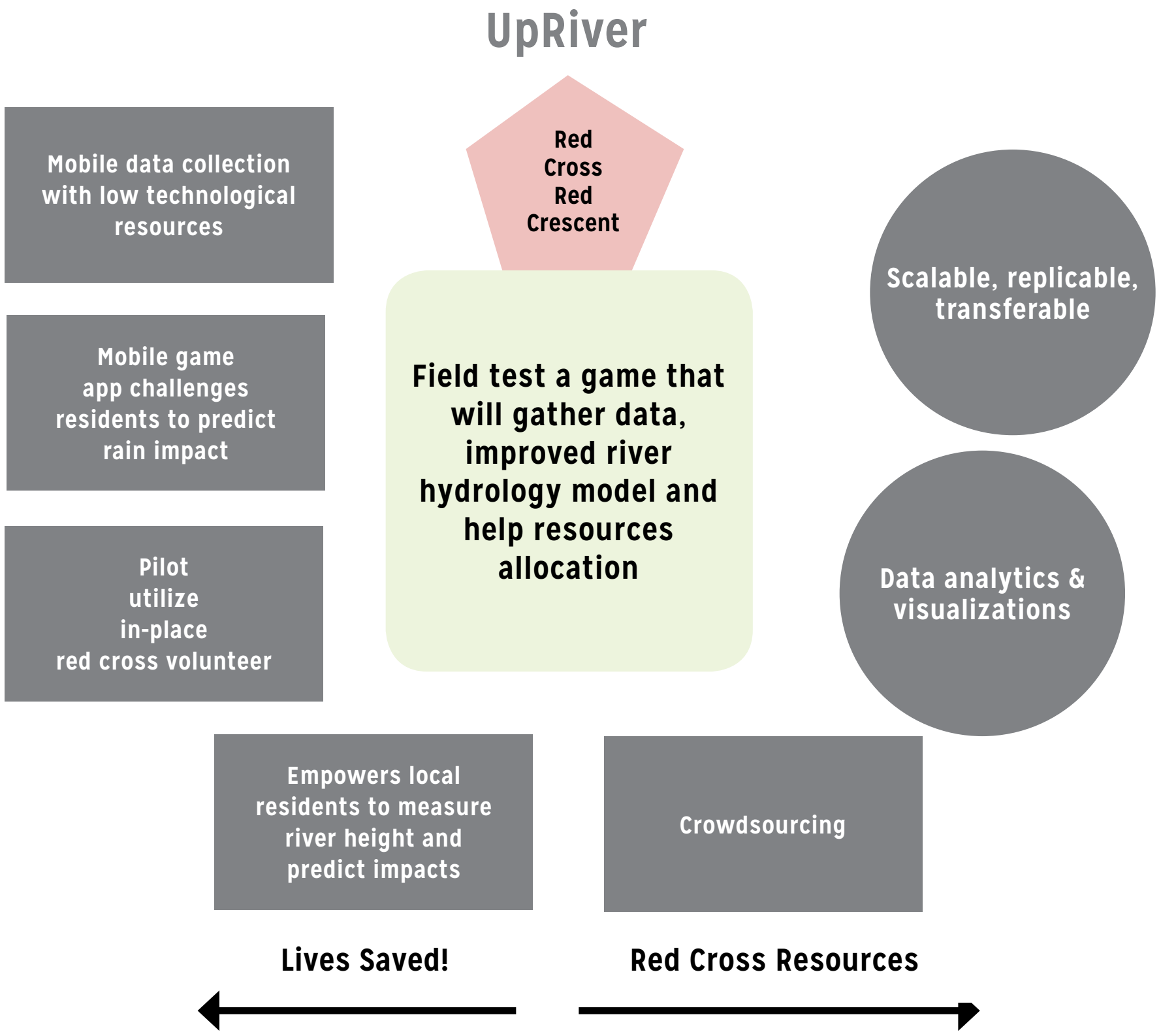

*Figure 6: UpRiver project diagram – local involvement in river level monitoring builds trust in early warning system*

At the end of the rainy season, the player who earned the most points within each community exchanges them for a prize – something of monetary value (cash, mobile phone credit, etc.). We believe this gives people incentives to read river gauges and engage in collecting and submitting data. This kind of game will simultaneously improve models and increase trust in those models, saving property and lives.

A suitable deployment venue would be the Climate Centre's promising flood risk management project in West Africa, in collaboration with the Togo Red Cross[2]. With proper support, the next steps would be to refine the game concept, deploy the game to the Mono River Basin in Togo, and add a research layer to investigate the efficacy of the proposed human computation approach. There are also opportunities to infuse human computation approaches to the development of the predictive hydrological model, for example distributing tasks to pre-calculate flood scenarios by running numerous simulations in a decentralized fashion.

---

[1] http://engagementgamelab.org/projects/upriver/

[2] http://www.climatecentre.org/site/news/456/togo-red-cross-prepared-for-preparedness

### Antisocial Computing

Social media and digital communication tools have largely been considered positive vehicles of change. However, the power of social media has been harnessed by extremists and terrorist groups to spread propaganda and influence mass thinking. As our government and corporations begin to rely more and more on social media and online crowdsourcing for situational awareness and data, they will need to be able to identify, track in real time, and mitigate the risks. Existing approaches to cyberthreat assessment and mitigation strategies overlook the societal aspect, which warrants the need for novel human computational methods.

A social network can be exploited to cause mayhem, ranging from cyber to physical attacks on individuals or corporations, to causing widespread social unrest or panic. The agents behind such exploits could be motivated by money (e.g., through extortion or market manipulation), antisocial tendencies, or they may be acting as agents of a terrorist group or an unfriendly nation state.

Setting up and conducting such an operation was a fundamental engineering problem (a field participants termed "disinformation engineering"), involving identifying desired outcomes, formulating strategies to achieve those outcomes, and then taking corrective measures when things don't go according to plan. It is only by understanding the technological innovations behind disinformation engineering that we can engineer protective technologies.

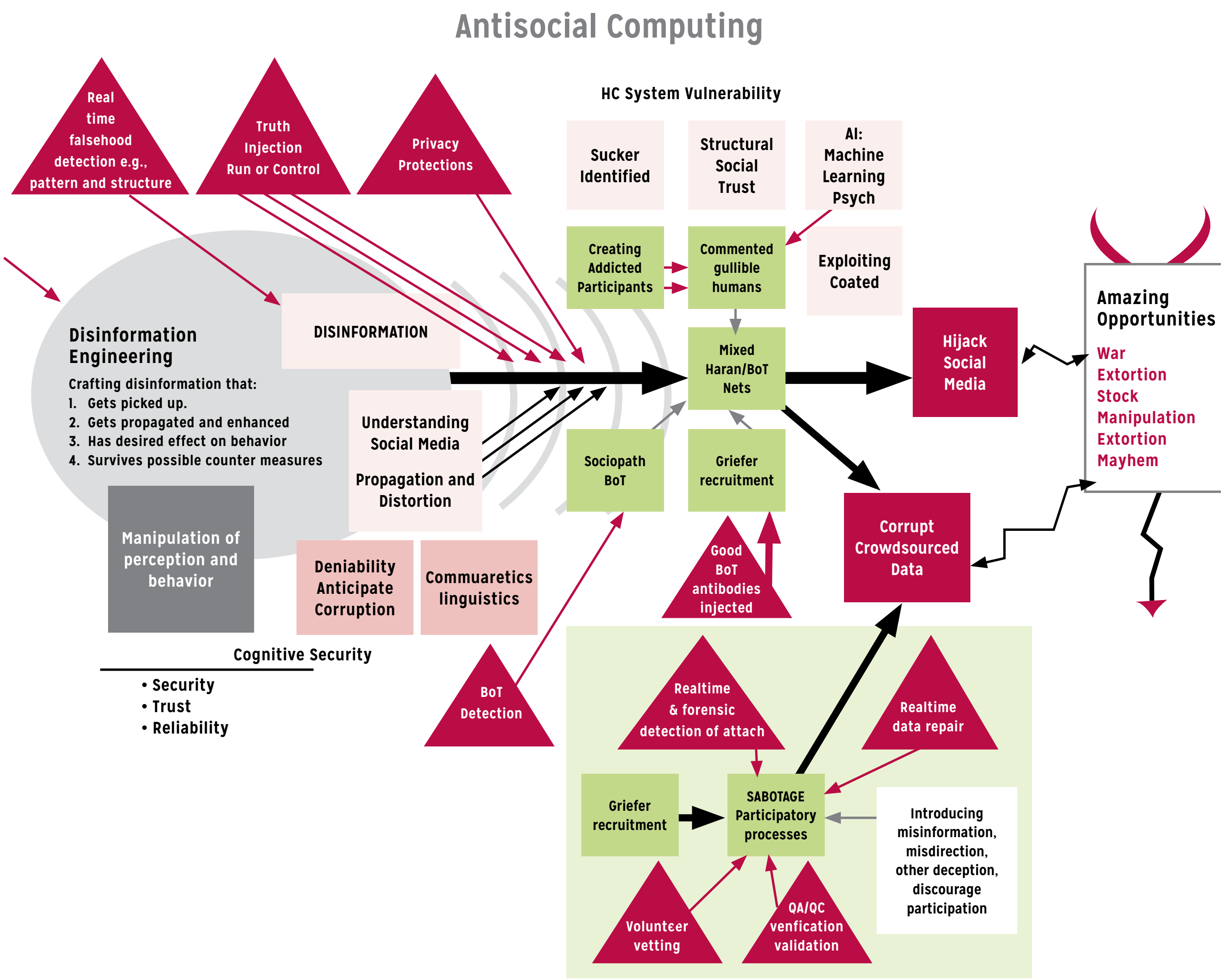

*Figure 7: Antisocial Computing project diagram – disinformation campaigns are enabled by HC technology*

Some elements of an attack would include:

- Preparing a social network that would enable cascading communication patterns that would rapidly amplify small seeds of disinformation on a global scale. This network could consist of gullible people (a.k.a. "suckers") inclined to believe the planted information, people inclined to cause trouble ("griefers") and computerized agents ("sociopathic bots").
- Manipulating the network through planted disinformation, dynamically steering the communications toward the desired outcomes. Here, the bots and griefers could serve as the active agents, while the suckers innocently propagate and corrupt the information they receive.
- Guarding against attempts by others to intervene (e.g., by planting truthful or counter information or by attempting to expose the agents and their conspirators).

Such a scheme would require careful design and an understanding of how people interact in social networks in order to manipulate those systems. Just as the adoption of networked information systems has led to entire new categories of disruption by cyber attack, human computation systems, especially social networks are already being exploited by malicious agents.

These ideas are further fleshed out in the article, "Antisocial computing: exploring design risks in social computing systems."[20]

## A New Set of Questions

These multi-day project explorations led to numerous useful insights about the new research challenges posed by human computation. When humans become part of the computational process, five new lines of inquiry arise: participation, analysis, architecture, design methods, and infrastructure.

### Participation

In contrast to the deterministic computing systems of today, humans have operating characteristics that vary from one individual to another. Moreover, human behavior is governed by a complex set of psychological and social phenomena. Therefore, the success of any system with humans in the loop depends heavily on a detailed and accurate understanding of factors related to participation. The following research questions seek to address this need:

- What are the ethical, legal, and social implications (ELSI) of human computation? What new issues arise in security, privacy, intellectual property, and fair labor practices and how should they be addressed? What are the roles, stakeholders, and power differentials that arise and how should we define best practices?
- How can systems be designed to be humanistic, that is, to ensure meaningful, dignified human participation? [22]
- What are the incentives that will attract and sustain a sufficient population of participants with the right skills to ensure a significant impact on the problem at hand?
- What are the most effective mappings between incentive models and project types to increase participation and effort?
- What are best practices in designing and governing a participatory system? For example: identify stakeholders, participant populations, a set of specific and overarching goals for the type of environment being developed, and then design to account for audience motivation and behaviors.
- How can methods be tailored and diffused to enable the poorest and most vulnerable sectors of the global population to engage in and benefit from human computation?

### Analysis

In many online systems, computation can be emergent rather than engineered[24]. In other words, information can arise through analysis as a useful byproduct of a large population of interacting individuals. For example, when people receive tweets as input in Twitter and produce tweets as output, the resultant activity traces can be analyzed in goal-directed ways, such as to predict events. The need for relevant analytic methods leads to the following questions:

- How can the mechanisms that underlie individual human behavior be revealed by online activity traces?
- How do such mechanisms inform models of collective behavior that arise from technology-mediated interactions among many individuals?
- What useful information and outcomes can be derived from collective behavior?
- What are best practices for measuring and classifying online social behavior for assessing its societal impact?
- How can the analysis of emergent collective behavior help inform the design of human computation systems?

### Architecture

Engineering new, effective human computation systems will require a conceptual framework for making high-level design decisions that address these questions:

- What classes of problems are most effectively addressed by human computation approaches? In other words, when is it appropriate to use human computation?
- Which architectural approaches are best suited to which problems (e.g., in crowdsourcing, sometimes we may wish to reassemble many individual human products into a single aggregate product, while in other cases, we may seek to identify the single best product)?
- What is the optimal division of labor between machines and humans that will result in a specific capability?
- How can machine capabilities be put to use for managing and evaluating the impacts of individual human variation?
- Given the variability of human behavior, what can we assert about the expected performance characteristics of the planned system? For example, how might we reliably estimate upper and lower performance bounds?

### Design Methods

Even with sound architectural principles in place, the core functionality of new human computation systems must be designed case by case. Both positive and negative examples of human computation design patterns currently exist. The following questions point to theoretical and empirical work that is needed to support repeatable methods that would ensure higher success rates.

- What are the basic project typologies, associated techniques, and interaction modalities?
- How do we design workflow architectures that most effectively combine human and machine input toward desired capabilities? How do we design to support emergent behaviors?
- system from malicious behavior? What is the potential impact on participants? How do we track such behavior in real time and what are effective countermeasures?
- How can expertise among participants be identified and leveraged?

### Infrastructure

The ability to answer the research questions above and create new human computation systems efficiently critically depends upon the existence and broad availability of specialized tools and network enhancements. The following research questions support the development of such human computation infrastructure:

- How do we build integrated software development environments (IDEs) that allow us to write, test, execute, and reuse code that operates on distributed human/machine systems?

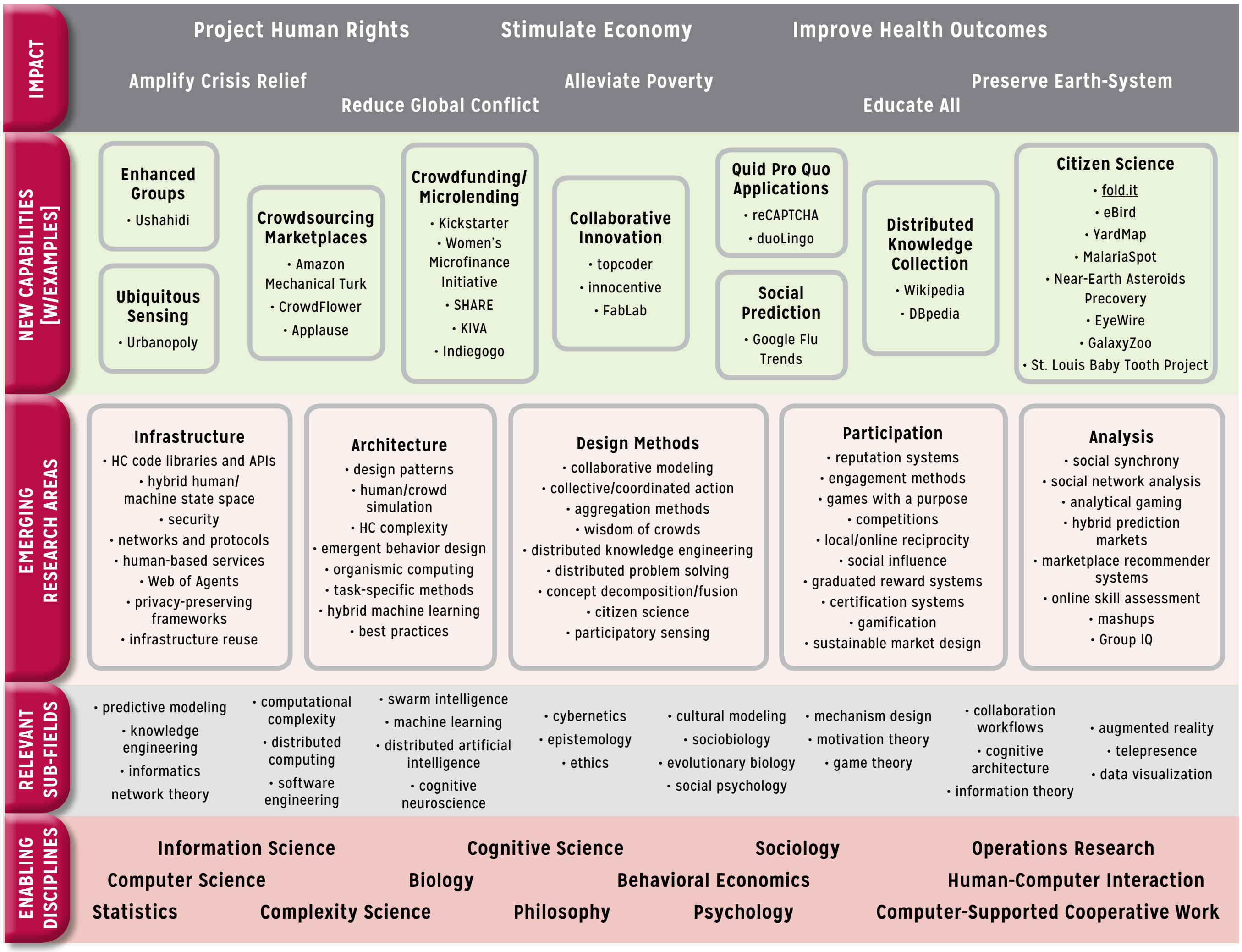

*Figure 8. High-impact societal benefits will be supported by a scaffolding of new human computation research.*

- How can we simulate human behavior in an IDE to reduce the financial and logistical costs of testing such systems before engaging potentially costly human participants?
- What embellishments are needed to the current infrastructure (e.g., Internet, communication protocols, etc.) that will enable always-on, asynchronous human participation?

## Emerging Research

A broad smattering of loosely connected research activities (the large yellow band in ***Figure 8***) has begun to address this list of research questions. However, most of these pursuits occur in isolation and ignorance of each other, due to their distinctive disciplinary origins (see "Enabling Disciplines" and "Relevant Sub-Fields" bands in ***Figure 8***) and consequent publication in narrowly scoped journals. Fortunately, these conditions are improving due to the open call to include other disciplines at the AAAI HCOMP conference and the new transdisciplinary journal *Human Computation*.

## Funding Environment

Despite these improvements in scientific communication, there is a paucity of U.S. federal funding for human computation research. The few counterexamples to this (e.g., the Cyber-Human Systems and Cyberlearning programs at the National Science Foundation) are notable for their pioneering vision. Furthermore, human computation, which often lies near the conceptual perimeter of established disciplines, is often at a disadvantage in competing for core program funding because there are relatively few qualified reviewers with sufficiently interdisciplinary backgrounds to evaluate the soundness of such proposals.

The situation is improving but not quickly enough. As a step in the right direction, the second recommendation of the 2013 President's Council of Advisors on Science and Technology (PCAST) report called for an interagency initiative to explore cross-agency collaboration in Social Computing [24]. This resulted in the formation of a Networking and Information Technology Research and Development (NITRD-SEW) subcommittee for Social Computing that brings together national funding agency representatives on a regular basis to learn more about ongoing work in this area. Fortunately, this seems to be engendering greater acceptance of the field and increasing awareness of related research. However, new funding programs have not yet directly resulted from this activity.

## National Initiative

We believe that the rapid advancement of this field toward repeatable and sustainable success models requires a concerted effort by policy-makers, federal funding agencies, multidisciplinary research institutions, private industry, and the public (via direct participation). Only through the collective action of these organizations and entities can we hope to endow human computation with the full apparatus of scientific inquiry and methodological maturity necessary to conscientiously[25] leverage the full transformative power of this new technology.

Furthermore, **we advocate the creation of a national center for human computation,** analogous to the National Socio-Environmental Synthesis Center (SESYNC), but dedicated to solving societal problems by bringing together different disciplines and stakeholders to develop human computation methods and capabilities. Due to the transdisciplinary nature of the field, we believe such a center would best support the rapid advancement of methods that might not be easily pursued in narrower contexts.

Toward these ends, **we propose a new national initiative in human computation,** with policy and funding support at all levels, to broaden and accelerate the research and development of collaborative information processing systems that leverage the respective strengths of machines and humans toward unprecedented capabilities to address our nation's and, indeed, humanity's most pressing societal needs.

## Contributors

### Organizing Committee

- Pietro Michelucci, chair, Executive Director, Human Computation Institute
- Janis Dickinson, Professor of Natural Resources, Cornell University; Director of Citizen Science, Cornell Lab of Ornithology
- Haym Hirsh, Dean of the Faculty of Computing and Information Science, Cornell University
- Lea Shanley, Former Director, Commons Lab of the Science and Technology Innovation Program, Woodrow Wilson International Center for Scholars

### Program Committee

- Matthew Blumberg, Executive Director, GridRepublic
- Michael Witbrock, VP for Research, Cycorp

### CCC Liaisons

- Ann Drobnis, Director, Computing Community Consortium
- Randal Bryant, Dean of the School of Computer Science, Carnegie Mellon University

### Policy Keynote

- Tom Kalil, Deputy Director for Technology and Innovation, Office of Science and Technology Policy, The White House

### Facilitators

- Luke Hohmann, Lead Facilitator, Sunni Brown Ink
- Stacy Weitzner, Live Visual Capture Facilitator, Sunni Brown Ink

### Workshop Support Staff

- Melissa Gedney, Woodrow Wilson International Center for Scholars
- Elizabeth Tyson, Woodrow Wilson International Center for Scholars
- Helen Vasaly Wright, Computing Community Consortium

### Complete List of Participants

- Dave Ackley, UNM CS
- Nitin Agarwal, University of Arkansas at Little Rock
- Larissa Albantakis, UW Madison
- Mary Catherine Bateson, George Mason University
- Nancy Baym, Microsoft Research
- Steven Becker, NIH
- Paul Bennett, Microsoft Research
- Andrew Bernat, CRA
- Abraham Bernstein, University of Zurich
- Jeff Bigham, Carnegie Mellon University
- Matthew Blumberg, GridRepublic
- Tom Boellstorff, University of California, Irvine
- David Brin, Science Fiction Novelist
- Sunni Brown, Sunni Brown Ink
- Randal Bryant, Carnegie Mellon University
- Edward Castronova, Indiana University
- Yiling Chen, Harvard University
- Adam Cheyer, Change.org
- Lydia Chilton, University of Washington
- Noshir Contractor, Northwestern University
- R. Jordan Crouser, MIT Lincoln Laboratory
- Kevin Crowston, National Science Foundation
- Janis Dickinson, Cornell
- Steven Dow, HCI Institute, Carnegie Mellon
- Edmund Durfee, University of Michigan
- Hamid Ekbia, Indiana University
- Christina Engelbart, Doug Engelbart Institute
- Michael Franklin, UC Berkeley
- Liane Gabora, University of British Columbia
- Susan Graham, University of California, Berkeley
- David Alan Grier, George Washington University
- Greg Hager, Johns Hopkins University
- Haym Hirsh, Cornell University
- Luke Hohmann, Sunni Brown Ink
- Panos Ipeirotis, New York University
- Tom Kalil, White House OSTP
- Geoff Kaufman, Tiltfactor Laboratory, Dartmouth College
- Markus Krause, Leibniz University
- Debbie Lockhart, NSF
- Stuart Lynn, Adler Planetarium / Zooniverse
- Thomas Malone, MIT
- Mimi McClure, NSF
- David McDonald, University of Washington
- Pietro Michelucci, Human Computation Institute
- Beth Mynatt, GT Institute for People and Technology
- Theodore (Ted) Pavlic, School of Life Sciences, Arizona State University
- Nathan Prestopnik, Ithaca College
- Lea Shanley, Science and Technology Innovation Program, Wilson Center
- Yiyang Shen, CRA
- Katie Shilton, University of Maryland College Park
- Paul Smaldino, Johns Hopkins University
- Nigel Snoad, Google Crisis Response
- Ram Sriram, NIST
- George Strawn, NITRD
- Pablo Suarez, Red Cross Red Crescent Climate Centre
- Seth Teicher, CrowdFlower
- Shailin Thomas, Berkman Center for Internet and Society, Harvard University
- Mike Webster, Macaulay Library, Cornell University
- Andrea Wiggins, DataONE
- Michael Witbrock, Cycorp Inc

# Appendix: The Visioning Process

### Overview

The goal of the three-day summit was to reveal research areas and opportunities that would lead to the design of human computation systems that generate high impact societal outcomes. Beginning with agreed-upon societal outcomes, we worked backwards to develop candidate solutions that would be enabled by human computation. The resulting catalog of such methods, in turn, pointed to the need for fundamental research in a specific set of domains. We hoped that mapping the fundamental research to new capabilities and outcomes would help inform a new national initiative leading to the anticipated societal benefits. Figure 9 depicts the intended three-day path for arriving ultimately at a comprehensive research roadmap that would connect fundamental research to new capabilities that address societal challenges.

We used participatory gaming, discourse, and introspective/collaborative analysis to examine motivation at different scales and stimulate innovation. This set the stage for collaborative analysis of additional material throughout the process.

### Shared Context

In this vein, Day 1 was kicked off with a simulation game conducted by participatory gaming expert Pablo Suarez of the Red Cross Red Crescent Climate Centre, which allowed workshop participants to directly experience aggregate and individual outcomes associated with participating in a human computation system, as well to observe their own responses to shifting incentive structures. The rest of the morning was a combination of presentations and lightning talks (see Figure 10).

These served both to set a historical context for discussing human computation and to get to know participants through the lens of their individual perspectives on the topic. Figure 11 is a "live visual capture" of highlights from the opening session, while Figure 12 depicts key concepts from participant presentations.

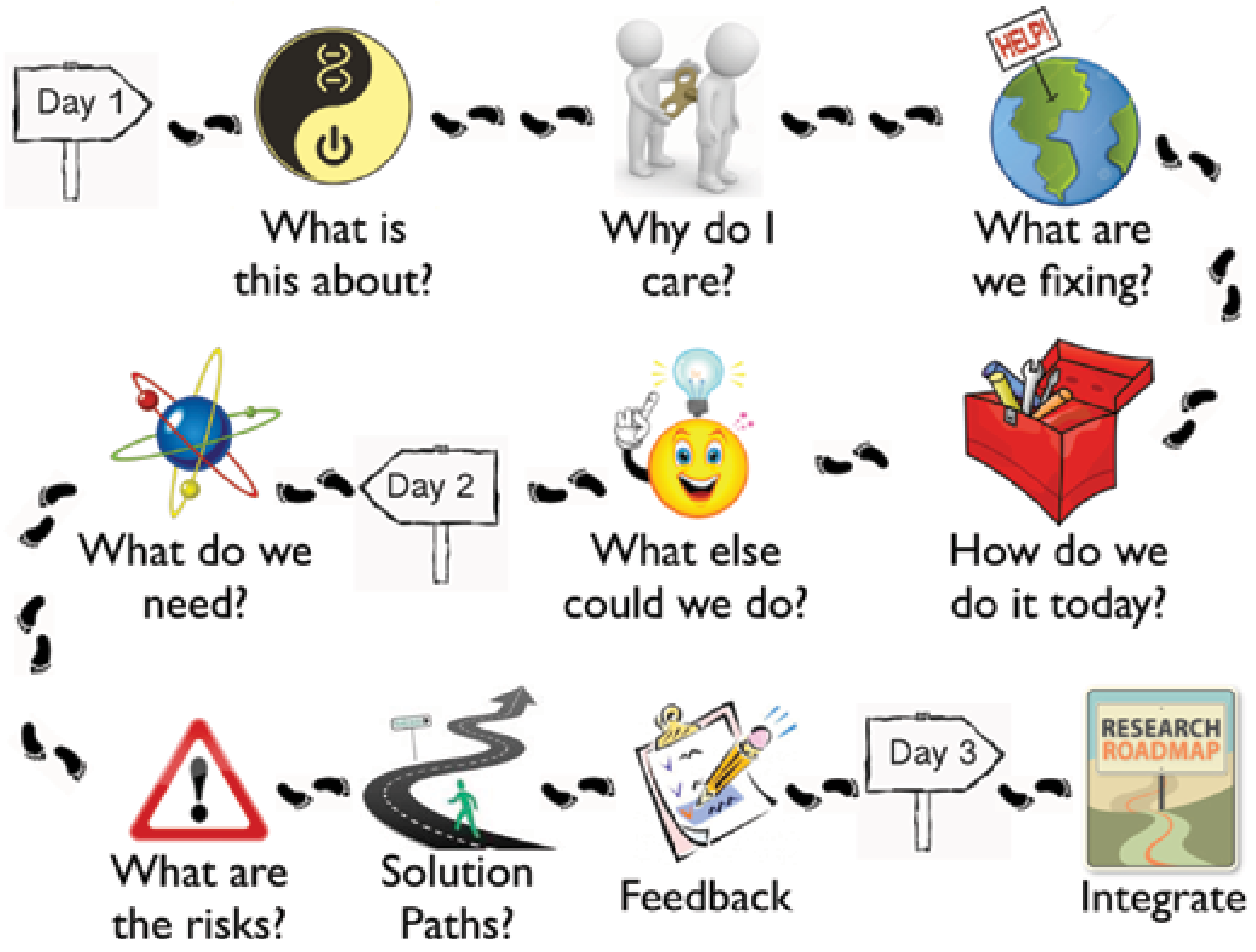

*Figure 9: Path of workshop activities leading to integrated research roadmap.*

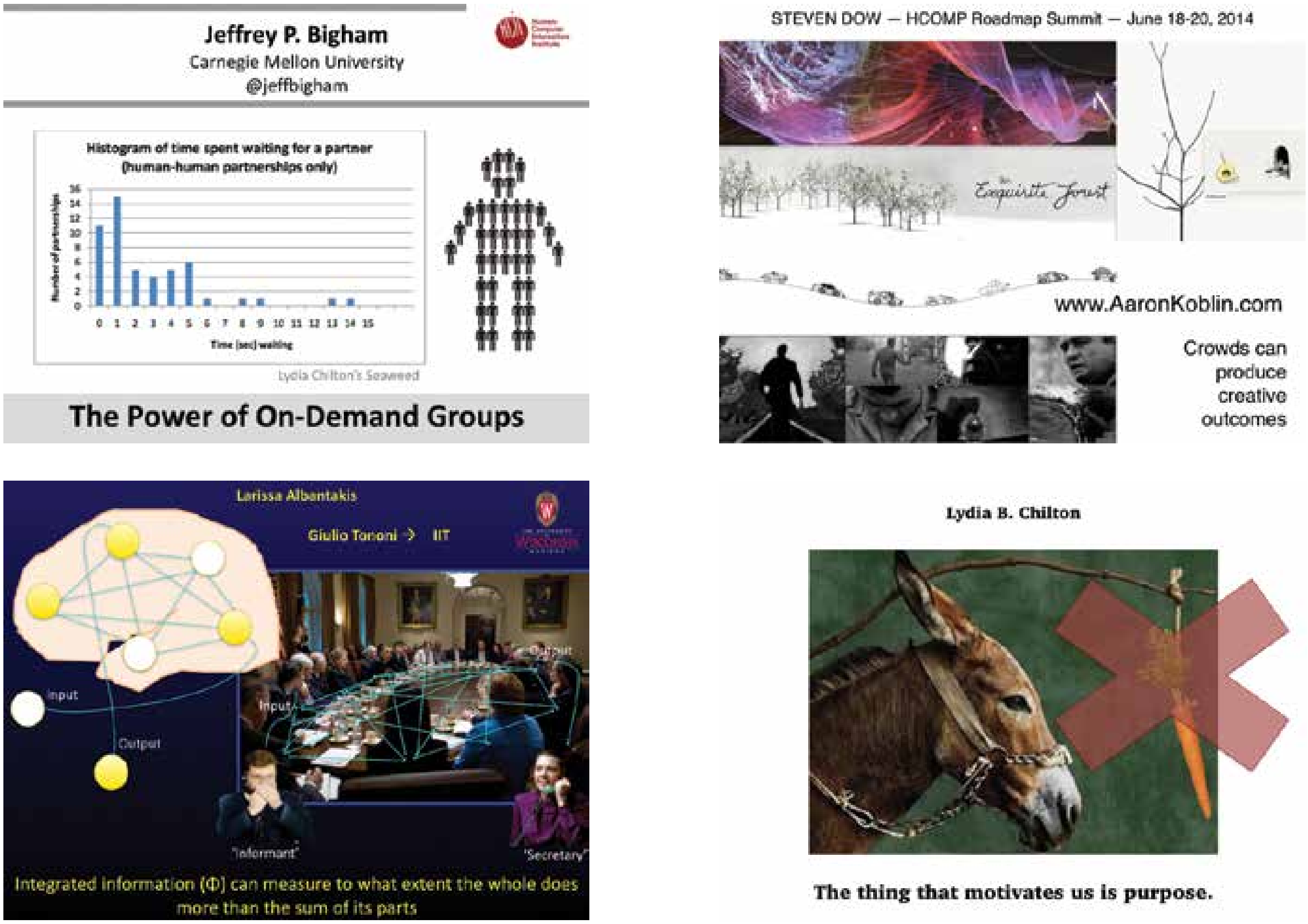

*Figure 10: Sample of four slides from lightning talks.*

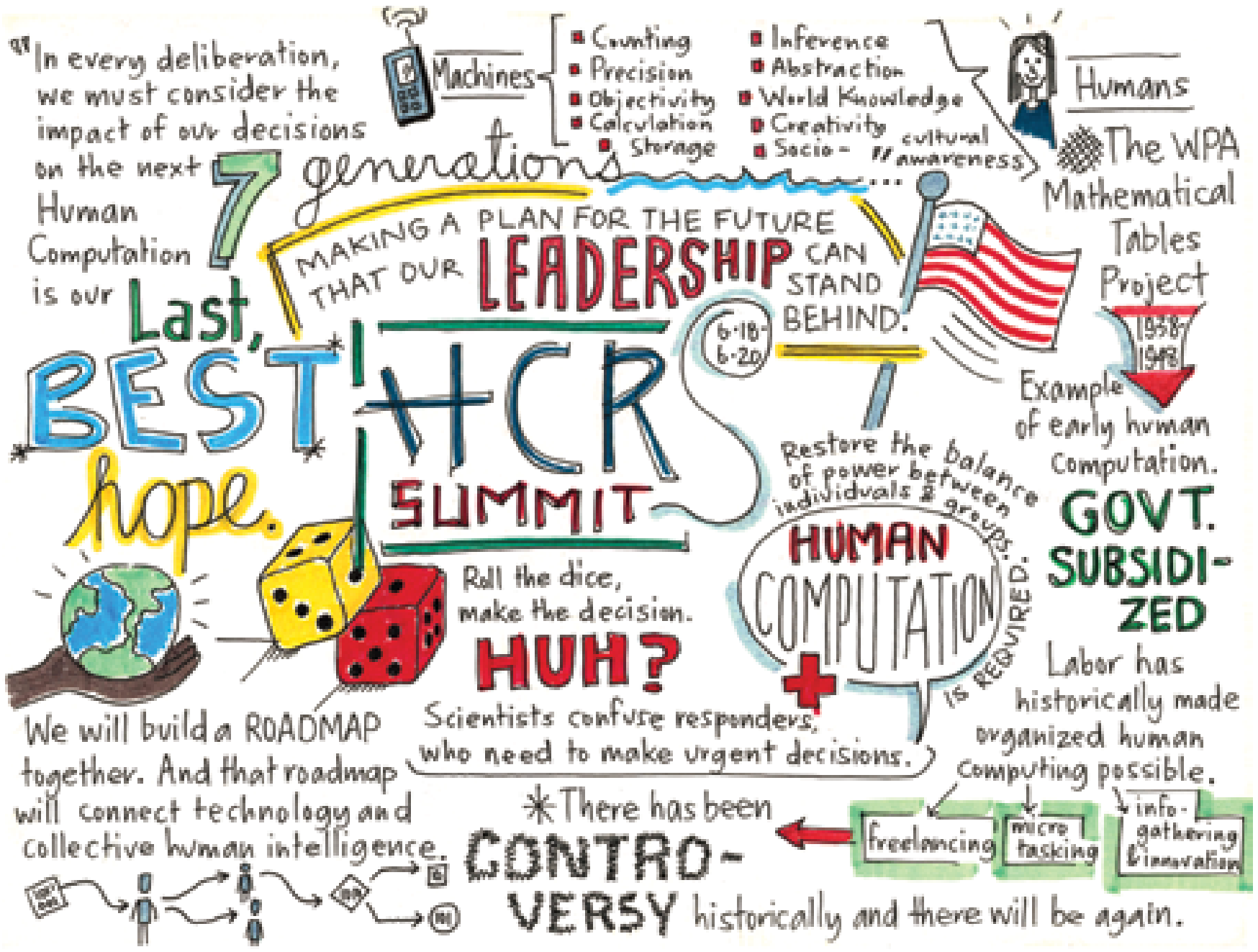

*Figure 11: Live visual capture of the opening session of the Human Computation Roadmap Summit.*

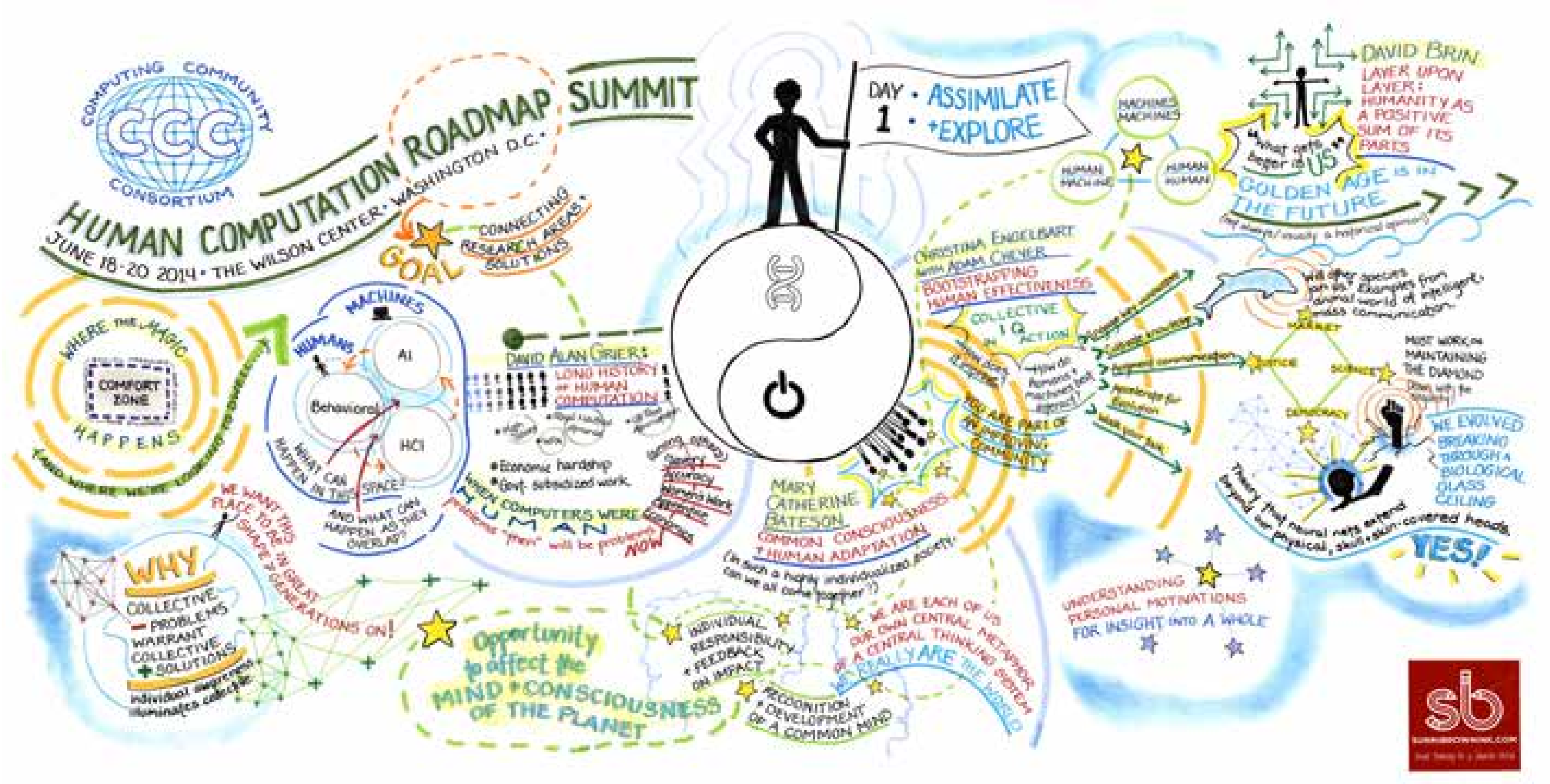

*Figure 12: Live visual capture of key concepts from invited speakers.*

### Problems and Solutions

The afternoon session of Day 1 was filled with participatory activities that generated introspection and sharing of personal motivations under the assumption that creative approaches arise out of meaning. This led to a multidimensional, generative analysis of societal problems ranging from somewhat tractable issues to wicked problems rife with uncertainties, feedbacks, and complexities that are poorly understood. This exercise, in turn, set a context for brainstorming about new human computation solution concepts. The live visual capture of this session's activities is illustrated in Figure 13.

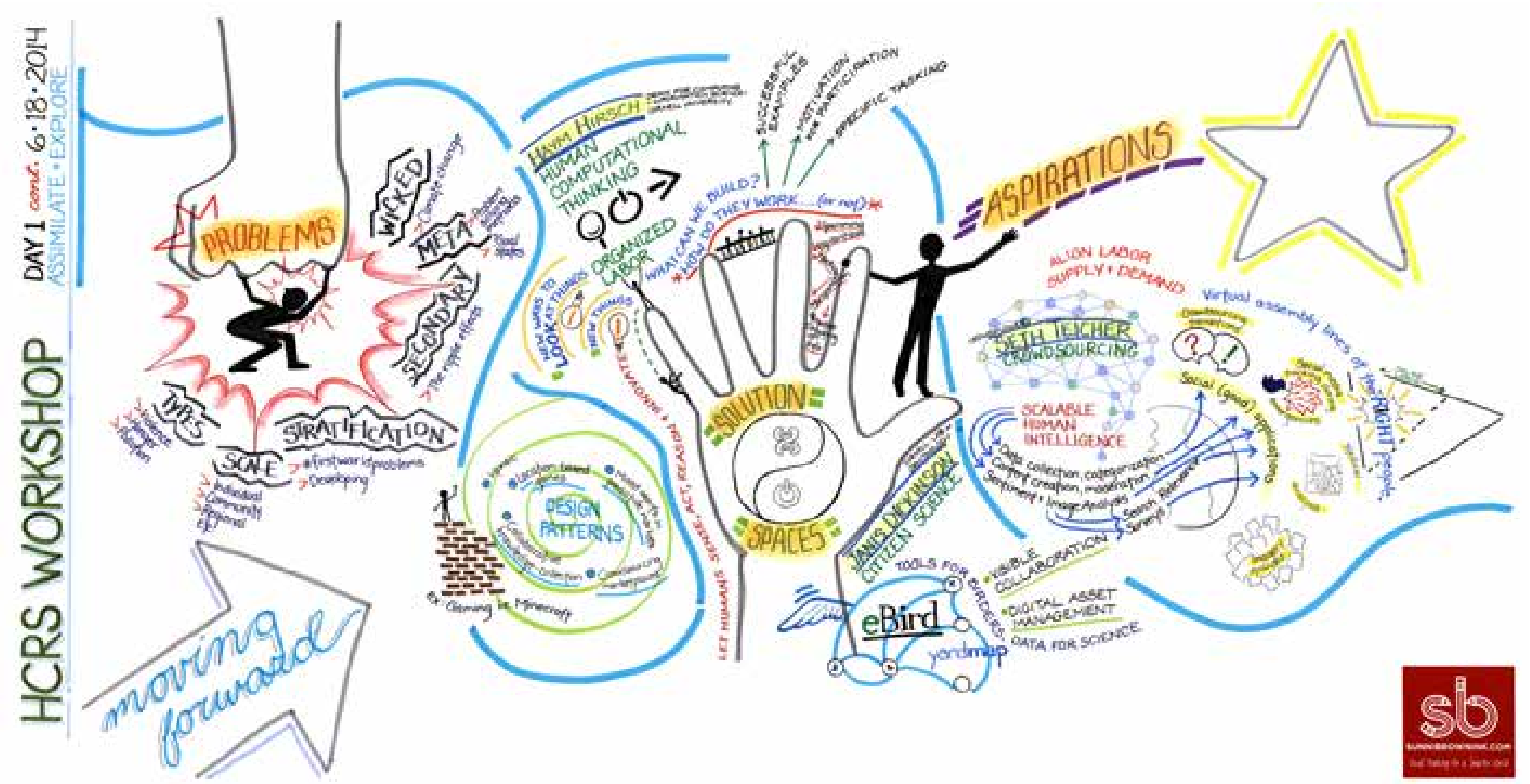

*Figure 13: Live visual capture of societal problem and solution generation activities.*

### Solution Concept Development

Day 2 began with a risk analysis exercise aimed at revealing the worst case scenarios associated with human computation system development. This was intended to bring awareness to the importance of incorporating risk analysis into emerging solution concepts. Next, participants were invited to aggregate around solutions of interest to form solution concept development teams. Eight solution teams self-organized to flesh out their human computation ideas (see Figure 14 and Figure 15) and prepared to present those ideas in the session that followed.

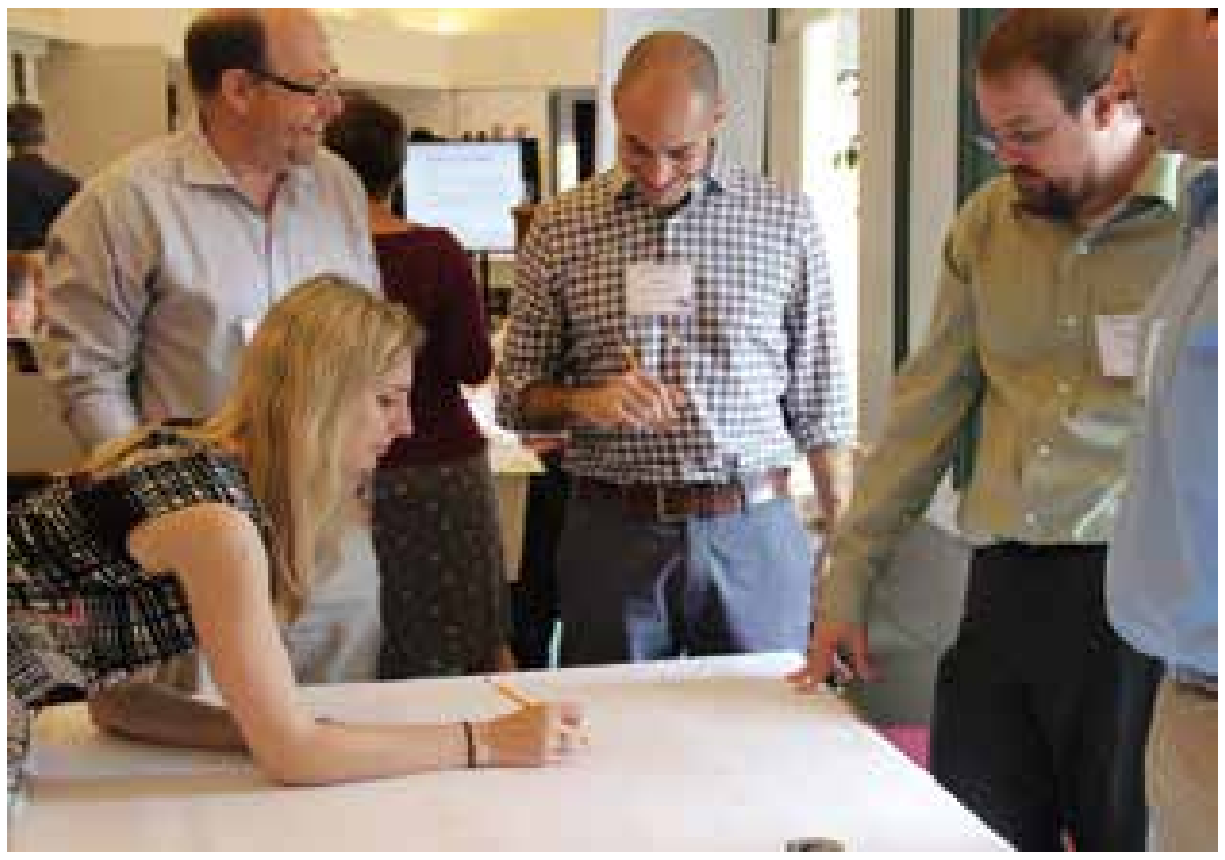

*Figure 14: A solution team begins to sketch out a concept.*

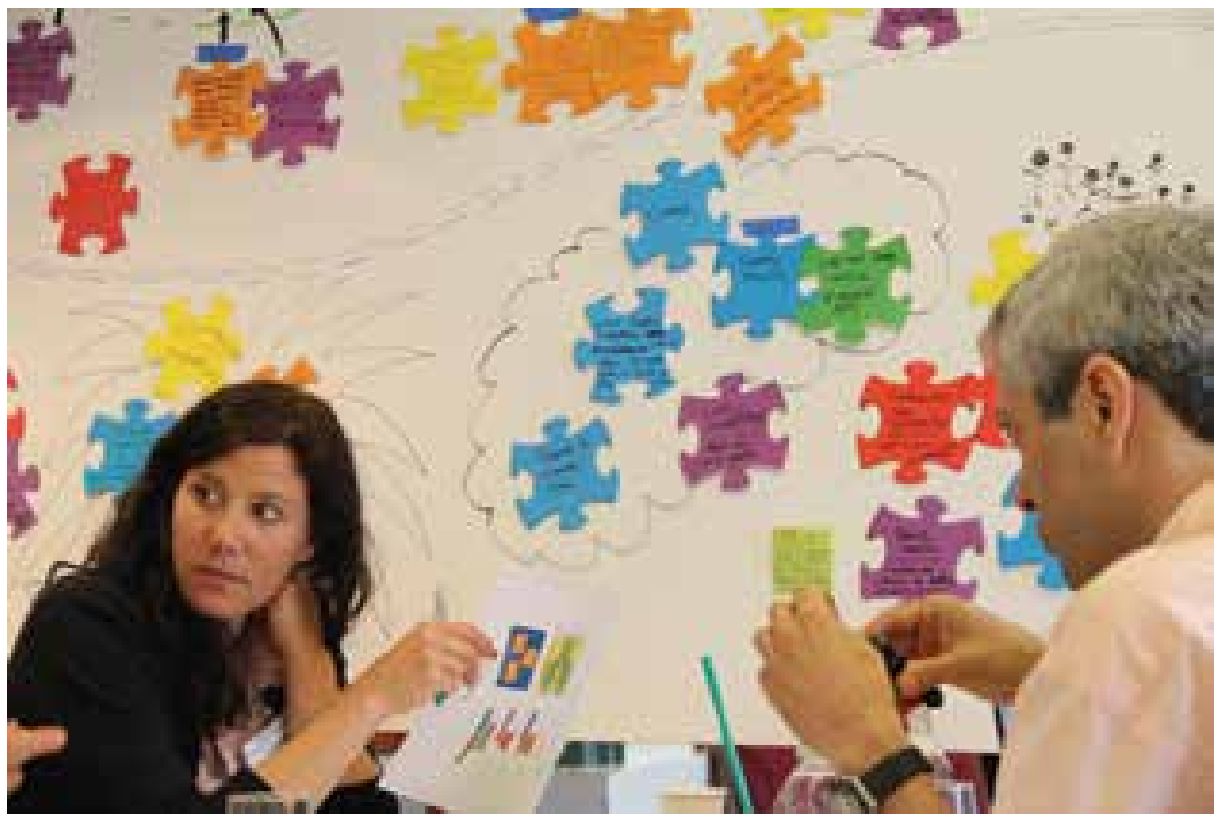

*Figure 15: A solution team prepares to present their concept.*

In the afternoon session, teams presented their concepts (Figure 16) to Tom Kalil (Figure 17), who provided constructive feedback through the lens of a policy-maker. This helped sharpen ideas and increased their accessibility to relevant audiences.

In addition to specific feedback about each solution concept, Tom Kalil provided general feedback by suggesting that the groups narrow the scope of the proposed solutions to more specific implementable projects. The activities of Day 2 are captured in Figure 19.

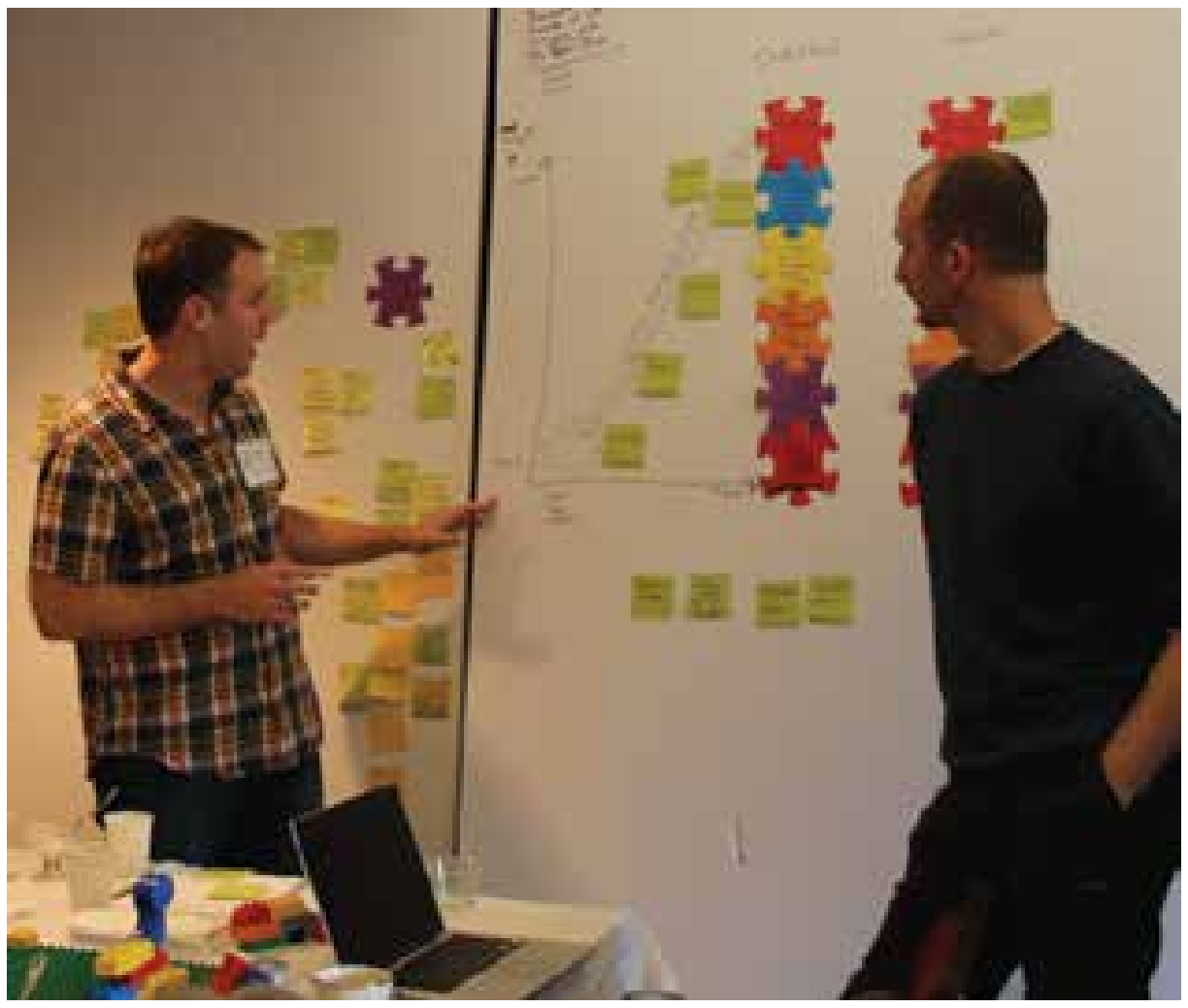

*Figure 16: A solution team presents its concept.*

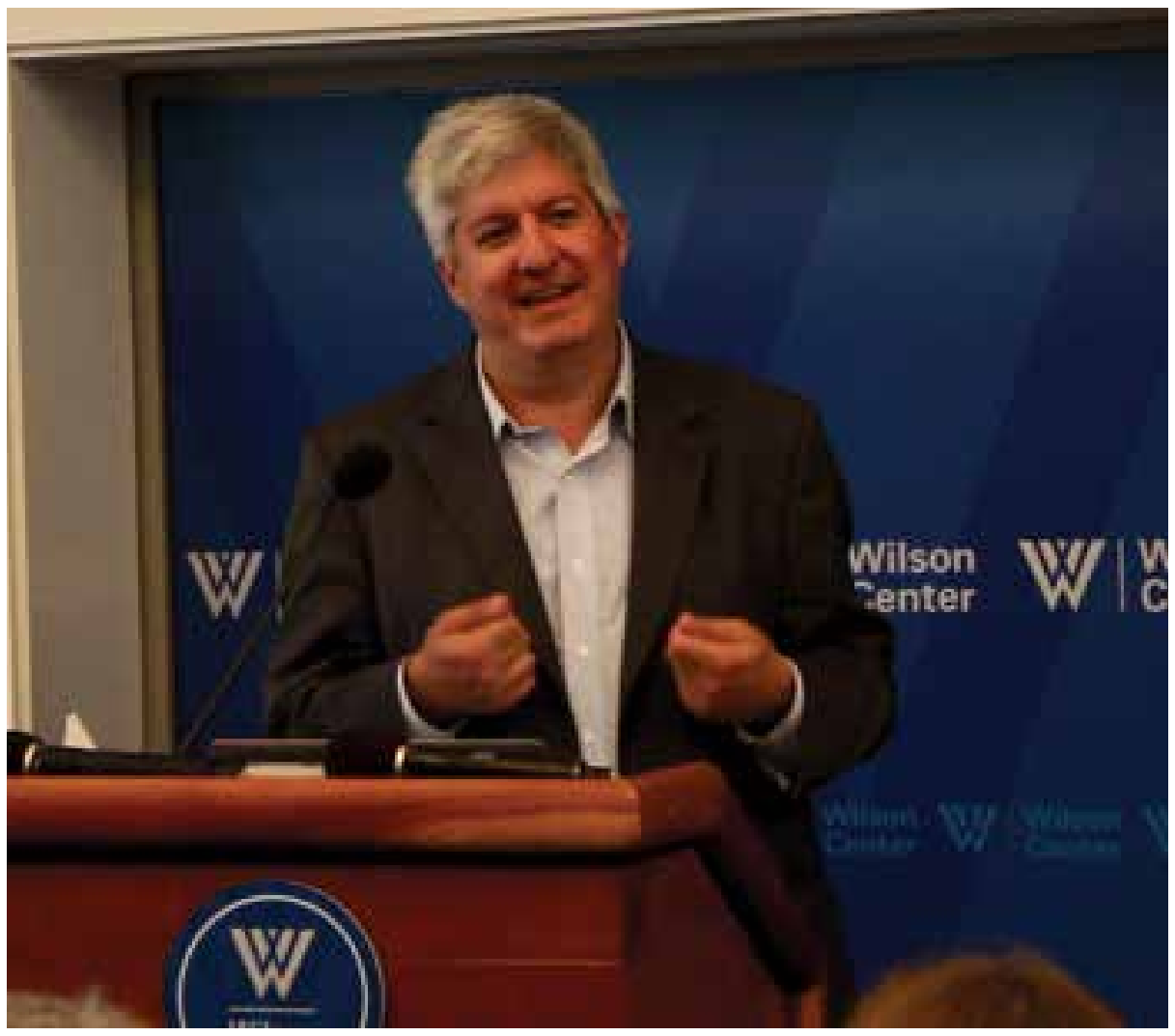

*Figure 17: Tom Kalil (Deputy Director for Technology and Innovation, White House Office of Science and Technology Policy) conveys a policy perspective on national initiatives.*

### Success Cases and Solution Refinement

Kalil's feedback suggested a deviation from the original plan for Day 3 activities. Originally, the third day of the workshop had been designated for integrating the new solution concepts into a single coalescent research roadmap for human computation. Instead, we responded to Kalil's feedback by using Day 3 to further refine the solution concepts and relegate the roadmap integration to post-workshop activities that would be left to the organizers. This detour is depicted in Figure 18.

Additionally, a decision was made to develop roadmap diagrams for recent human computation success cases

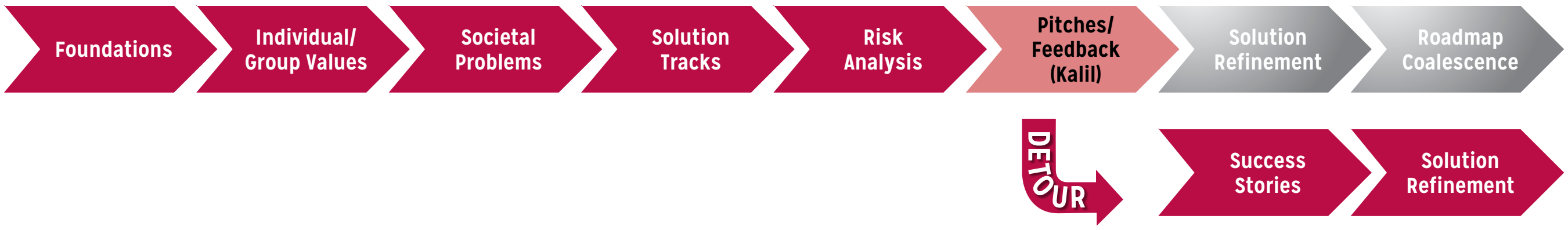

*Figure 18: Kalil's feedback suggested a course-correction for Day 3 of the workshop to further concretize the solution tracks.*

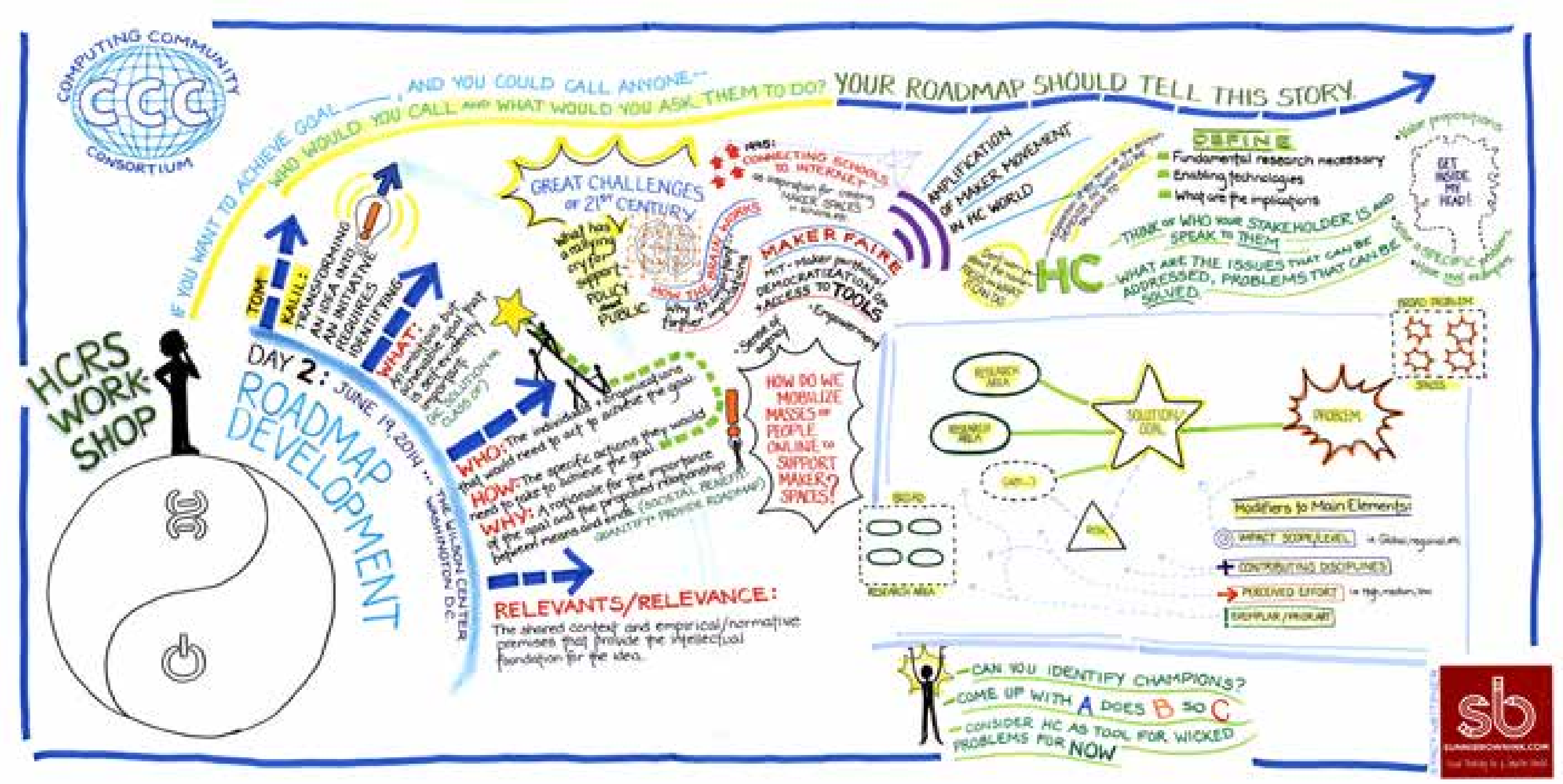

*Figure 19: Live visual capture of Day 2: Individual solution generation and pitch feedback from Kalil.*

(Figure 20) prior to developing such diagrams for the workshop-generated solutions. This served two purposes: 1) a deep technical assessment of the success cases helped inform a down-select of the workshop-generated solutions to those most amenable to the recommended scoping, and 2) the full set of solution diagrams, including both old and new, would help ensure a more complete and representative assessment of the underlying research space.

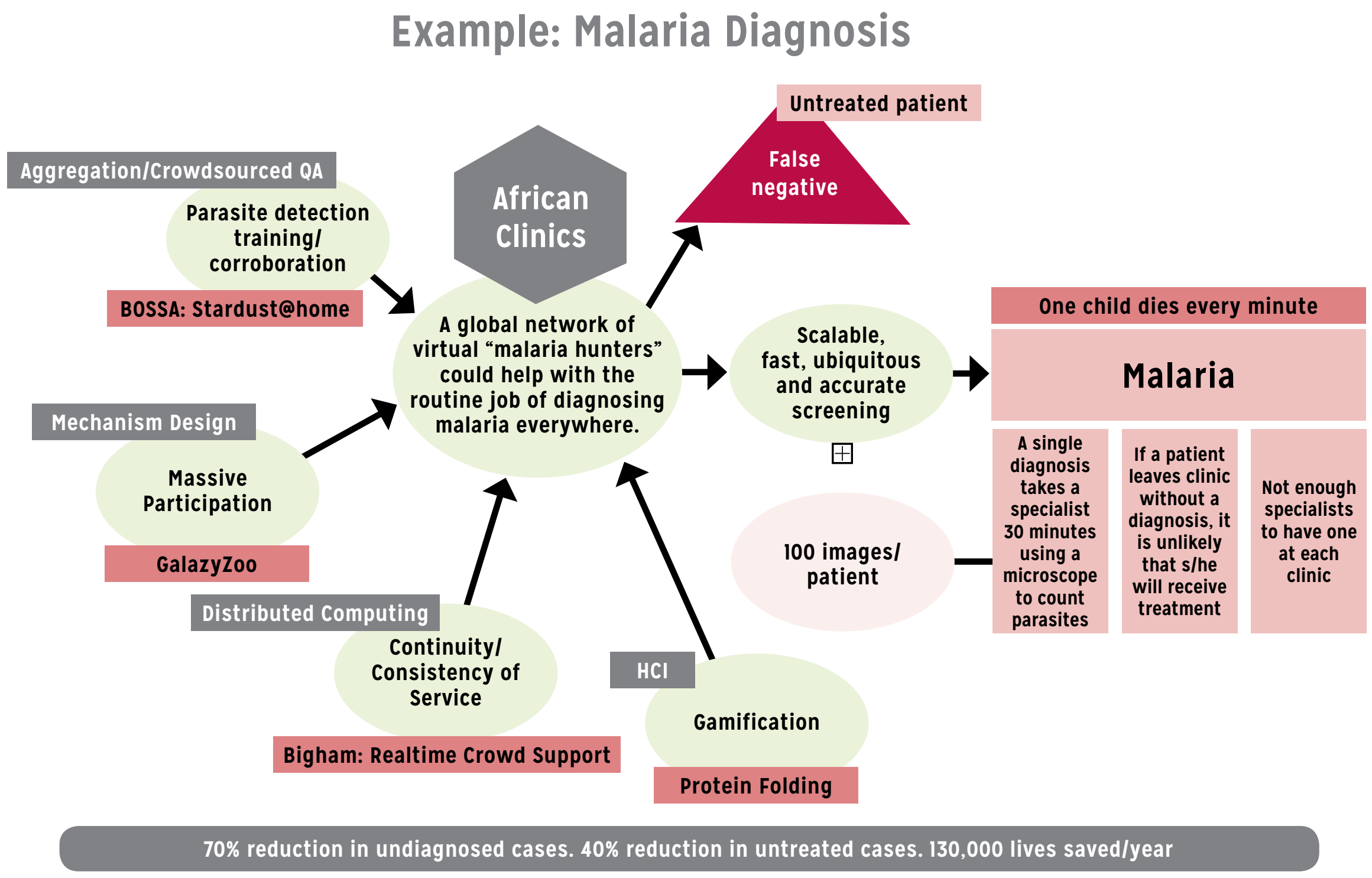

*Figure 20: Roadmap diagram for human computation success story.*

Pablo Suarez kicked off Day 3 with a fast-paced participatory game called "Snap!", which might be described as an offline, concept-based implementation of the "ESP game"[26], designed to achieve some measure of descriptiveness and consensus around the term "Human Computation". The results of this exercise are presented in Figure 21 as a word cloud.

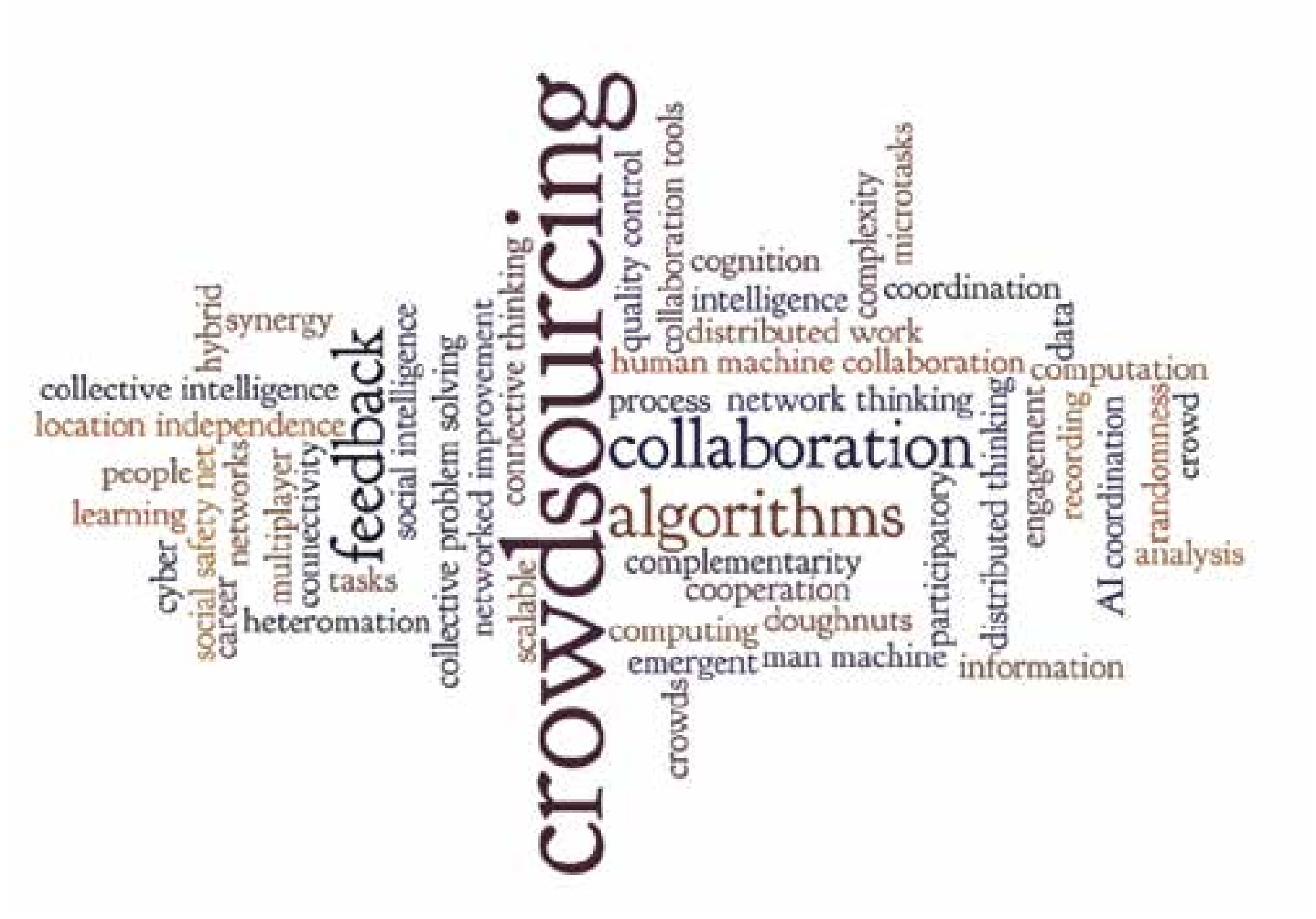

*Figure 21: Human Computation word cloud.*

Once the roadmap diagrams were produced for both historical success cases and the workshop-generated ideas (e.g., Figure 22), the workshop concluded.

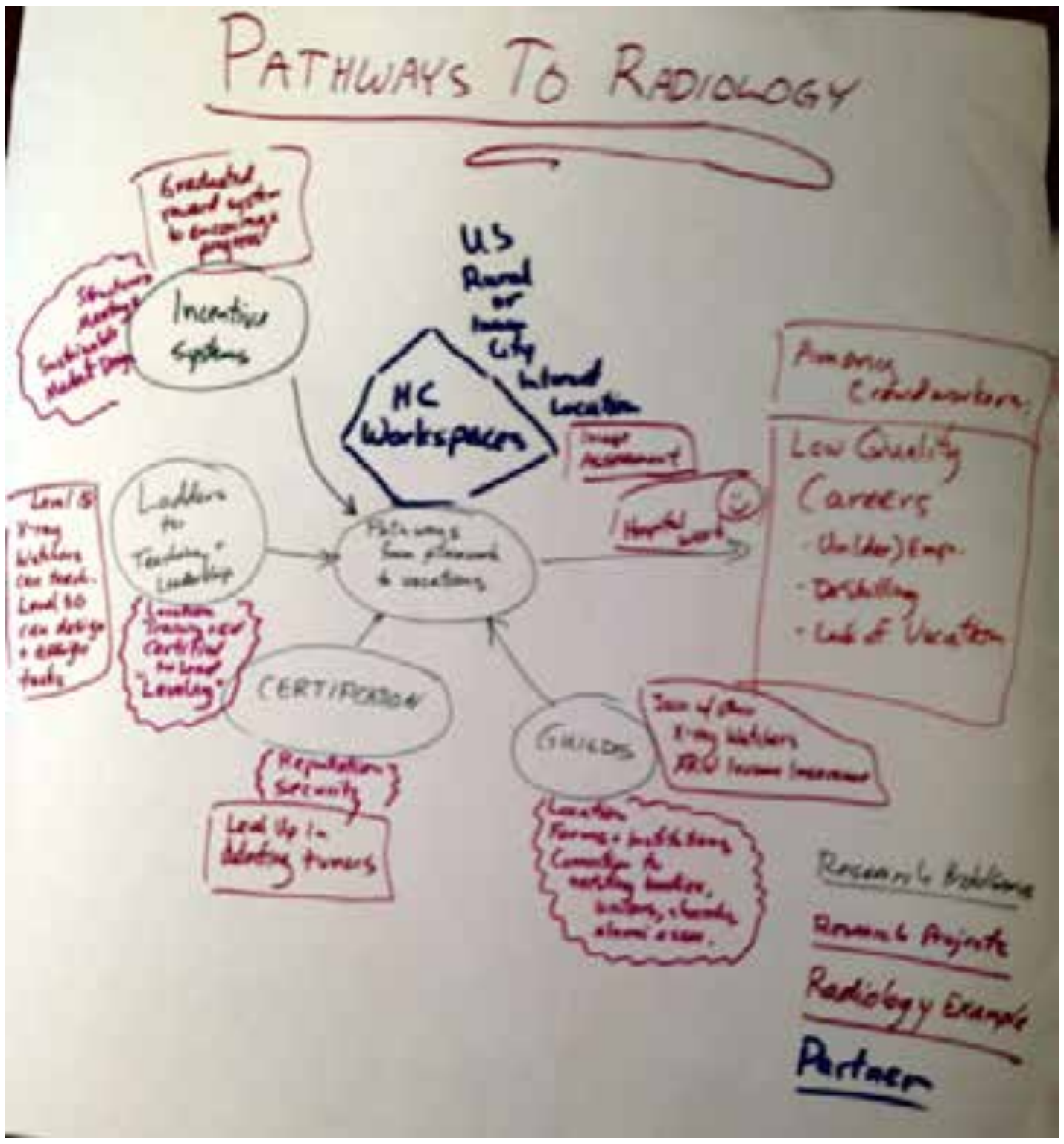

Figure 22: Roadmap diagram for new human computation solution.

The success cases and new ideas that were selected and diagrammed in Day 3 are visually summarized in Figure 23. However, the new project ideas are detailed in the body of this report.

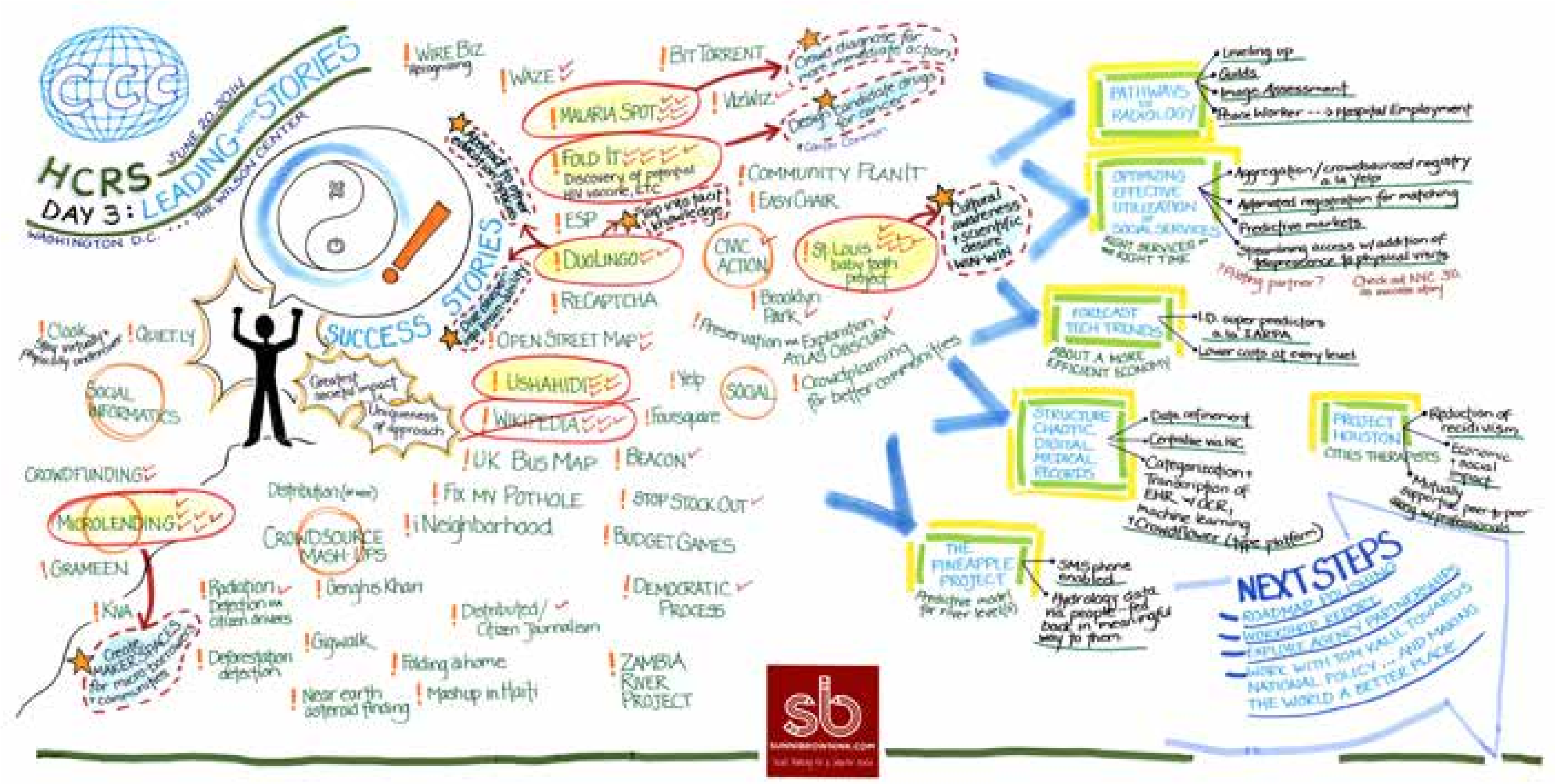

Figure 23: Live visual capture of historical human computation success stories and new human computation solution ideas.

## Notes:

CCC
Computing Community Consortium
Catalyst

1828 L Street, NW, Suite 800
Washington, DC 20036
P: 202 234 2111 F: 202 667 1066
www.cra.org cccinfo@cra.org